\documentclass[reprint,amsmath,amssymb,aps]{revtex4-2}
\usepackage{graphicx}
\usepackage{hyperref}
\usepackage{xcolor}

\usepackage{color}
\usepackage{amsmath}
\usepackage{mathrsfs}
\usepackage{amsfonts}
\usepackage{amssymb}
\usepackage{varioref}
\usepackage{float}
\usepackage{dcolumn} 
\usepackage{bm}
\usepackage{relsize}
\usepackage[flushleft]{threeparttable}
\newcommand{\cblue}{\textcolor{black}}

\usepackage{epstopdf}

\begin{document}
\title{The explicit characterization of counterion  dynamics \\ around a flexible polyelectrolyte}
\author{Keerthi Radhakrishnan}
\email{keerthirk@iiserb.ac.in}
\author{Sunil P Singh}
\email{spsingh@iiserb.ac.in}
\affiliation{Department of Physics, Indian Institute Of Science Education and Research, \\ Bhopal 462 066, Madhya Pradesh, India}

\begin{abstract} The article presents a comprehensive study of counterion dynamics around a generic linear polyelectrolyte (PE) chain with the help of coarse-grained computer simulations. The ion-chain coupling is discussed in the form of binding time, mean-square-displacement (MSD) relative to the chain, local ion transport coefficient, and spatio- temporal correlations in the effective charge. We have shown that a counterion exhibits sub-diffusive behavior 
$\langle \delta R^2 \rangle \sim t^\delta$, $\delta\approx0.9$ w.r.t. chain's centre of mass (COM). The MSD of ions perpendicularly outwards from the chain segment exhibits a smaller sub-diffusive exponent compared to the one relative to the chain's COM.
 Further, we confirm that the effective diffusion-coefficient of counterions is strongly coupled with the chain. The effective diffusivity of ion is the lowest in chain's close proximity, extending up to length-scale of the radius of gyration $R_g$. Beyond $R_g$ at larger distances, they attain diffusivity of free ion with a smooth cross-over from the adsorbed regime to the free ion regime. 
We have shown that the effective diffusivity drastically decreases for the higher valent ions, while the crossover length scale remains the same.   Conversely, with increasing salt concentration the coupling-length scale reduces, while the diffusivity remains unaltered. The effective diffusivity of adsorbed-ion reveals an exponential reduction with electrostatic interaction strength. We further corroborate this from the binding time of ions on the chain, which also grows exponentially with the coupling strength of the ion-polymer duo.   Moreover, the binding time of ions exhibits a weak dependence with salt concentration for the monovalent salt, while for higher valent salts the binding time decreases dramatically with concentration. Our work also elucidates fluctuations in the effective charge per site, where it exhibits strong negative correlations at short length-scales. 
\end{abstract}

\maketitle
\section{Introduction}

 Polyelectrolytes of biological origin such as  DNA, RNA, proteins, or other synthetic ones are ubiquitous in nature and their contribution to the evolution of life,  biological functions, and industrial applications are indispensable \cite{muthukumar201750th,Muthu_Book_2011,khokhlov1982theory,dobrynin1995scaling,holm2004polyelectrolyte,netz2003neutral,barrat1997theory}.  Typically,  a PE chain constitutes of ionizable groups attached to its backbone, which when immersed in an aqueous solution  dissociates, leaving behind charge on the chain's surface along with oppositely charged free counterions in the solution\cite{Oosawa_1971}. The incubation of long-range electrostatic interactions in the makeup of these polymeric systems in the form of charges pave way for strong correlations among the free ions and chain at various time-and length-scales\cite{Winkler_PRL_1998,muthukumar2006simulation,Grass_PRL_2008,hsiao2006salt,kundagrami2008theory,lo2008dynamical,dobrynin1995scaling,radhakrishnan2021collapse}.  Therefore, the size, charge, structure, and even the dynamical behavior of the PEs are coupled with motion of these small ions in an extremely non-linear fashion. {\cblue As a consequence of that the PE solution exhibits  multitudes of relaxation modes spread over various length-scales, as observed in the dielectric spectroscopy experiments\cite{bordi2004dielectric}. } Further, physiological factors associated with this Coulombic mixture such as the concentration, size and valency of the added salt play a crucial role in dictating the macroscopic structure of the PE\cite{hsiao2006salt,Liu_JCP_2002,golestanian1999collapse,barrat1997theory,Keerthi_Singh,radhakrishnan2021collapse}. The counterion effectuated phenomenologies in PE encompasses: coil to globule transition \cite{Winkler_PRL_1998,Muthu_JCP_1997,Muthu_JCP_1997}, chain collapse under multi-valent salt\cite{liu2003polyelectrolyte} and subsequent re-expansion\cite{hsiao2006salt} at higher concentrations, bundle formation in semi-flexible chains\cite{golestanian1999collapse}, etc\cite{gonzalez1995ion} . 


For any macromolecule, the ion atmosphere constitutes that region around the molecule, where the density profile of the free ions is different from the  bulk, due to its electrostatic couling with the polyion. The theory suggests\cite{manning1979counterion,fixman1980charged}, the surrounding electric double layer (EDL) essentially consists of a monolayer of condensed ions (\textit{Stern layer}) inducing an effective charge, followed by a distribution (\textit{diffuse layer}) decaying exponentially into the solution due to screening from the condensed ions. The ion dynamics within these layers largely influence the polyions properties like its relaxation time\cite{fixman1980charged,rau1981polarization,washizu2006electric}, dielectric response\cite{katsumoto2007dielectric}, conductivity\cite{fischer2008salt,Winkler_JCP_2009,Grass_PRL_2008}, etc. 
Further, for flexible or semiflexible PEs, the charge regularization is strongly dictated by the chain's conformational degree of freedom\cite{muthukumar2004theory, liu2002langevin}. This leads to strong ion-chain correlations at various length scales, extending up to macroscopic scales of chain dimension. 

The aim of the present work is to provide a detailed study of the ion-chain coupling, in terms of different modes of ion transport. The crucial quantity used for the characterization of ions in this context is the diffusivity of ions, as it is closely related with the mobility, ionic conductivity and dielectric properties\cite{bordi2004dielectric,katsumoto2007dielectric,fixman1980charged,smiatek2020theoretical} of the  PE. The focus of most of the past studies addressing ion diffusivity  has been on how at sufficiently large times counterions exhibit suppressed diffusivity\cite{Grass_PRL_2008,karatasos2009dynamics} and mobility\cite{Grass_PRL_2008,Frank_EPL_2008,Singh_2014_JCP}, resulting in diffusional retardation in contrast to the free ions. However, a crucial aspect of looking at the ion’s movement w.r.t. the chain remains unprecedented, which might lead to surfacing of distinct modes of ion transport. Interestingly, our findings suggest that the short-scale effective diffusivity exhibited by ions manifest a smooth continuous dependence over the chain-ion spatial separation. Thereby, it serves as an optimum tool in  demarcating various regimes of chain-ion coupling within the bounds of adsorbed and free ions.  Another important feature of the short scale diffusivity of ions cohabiting the chain is that it is nearly independent of changing proportions\cite{chang2002brownian} of the system like  molecular length,  concentration of counterions or salt ions, etc. However, it depends on those physiological parameters that defines the interactive strength like solvent quality, temperature, or valency in presence of salt. 


This approach is especially conducive in obtaining different modes of diffusion, where a suppressed diffusivity is found perpendicularly outwards from the chain backbone compared to the intersite hopping\cite{cui2007counterion,lo2008dynamical} along the chain, indicating longer residence time of ions. 
In purview of this, a section of the current work addresses the association time of ions with the PE chain. Apart from ion's diffusive timescale, the binding time  (spanning nanoscales) serves as another important timescale exhibiting profound effect on the chain's dynamics as reported in  previous studies\cite{cui2007counterion,prabhu2004counterion,morfin2004adsorption}. Following this influence of ion's association time-and length-scales on binding mechanism of ligands and proteins in real experiments, the past few decades, has seen a constant quest  to explicitly characterize and visualize ions dynamics  around nucleic acids and proteins\cite{prabhu2004counterion, yu2021experimental,shi2021counterion,schipper1997counterion}. But the dynamic nature of site binding of these ions, accompanied by rugged surfaces of the biomolecules and other aspects pose considerable challenges even in high resolution scattering experiments\cite{yu2021experimental}. Our explicit characterization of ion flux near a generic PE chain, unveils interesting facets such as exponential growth of binding time with Bjerrum length, non-dependence of free time, further supplemented by salt ions 
where the binding time exhibits uncanny dependence on ion concentration near the chain. 

 Further, an important ramification of the charge regularisation along the chain backbone is the effective charge imparted to the chain that is a precursor to
most of the phenomenologies exhibited by charged macromolecules\cite{torres2017protonation,mason2008protein, lund2005charge, lund2013charge}.  While most of these studies take into account the chain ionization in a global manner, overleaping local fluctuations, we perceive the PE in its innate form of inhomogenously sequenced array of ionized and non-ionized repeat units dynamically varying over time. Our results elucidate strong short scale spatial correlations in the  charge fluctuations per site. Although, such short scale collective correlations in counterion excitation near chain has been substantiated previously\cite{karatasos2009dynamics, Koplik_CFP_1995}, its manifestation in the form of chain's segmental charge fluctuations hitherto remains unaddressed. More so the pronounced effect of these short scale charge fluctuations have been previously reported to be pivotal in many multi-body complexation and binding phenomenons\cite{angelini2005structure, kirkwood1952forces,blanco2019role,da2009polyelectrolyte,lund2013charge}.

This article essentially entails a microscopic characterization of the ion dynamics within the counterion cloud surrounding a generic flexible PE chain using extensive molecular dynamic simulations. Explicit solvent effects with long-range HI are included.
{\cblue Here, molecular interactions involving solvent and other chemical specificities at further microscopic levels are not considered. The inclusion of these addendums at the cost of analytical and computational complexity is more important when dealing with specific systems\cite{fahrenberger2015influence,smiatek2020theoretical}.}
We undertake an elaborate study on the distinct modes of ion diffusion, consisting ion transport w.r.t to the chain COM, perpendicularly outwards from the backbone,  and its associated time-and length-scales. Further, dependence on the relevant physiological conditions like solvent quality, molecular length, ion concentration, etc. are also addressed. Moreover, the diffusive scales  are complimented with the quantification of ion association times of adsorption and desorption across the Stern layer. An extension of all these analysis is done for the case of added salt, where the collective ion excitations show interesting dependencies on salt concentration and valency. 



The article can be read as following: Section II  presents the  simulation model, and Section III, IV, V, VI, and VII entails the results. Section III presents the ion distribution, section IV discusses the dynamics of ions and effective diffusivity, and section V addresses binding time of ions. The  correlations of effective charge are shown in section VI, while the effect of salt concentration and  valency are presented in section VII.  The discussion and conclusion of the results are presented in  Section VIII.

\section{ Simulation Model } 
Our model primarily consists of a charged  linear polymer immersed in a neutral background  solvent consisting explicit counterions and  added salt-ions of different valencies.
The PE is modelled as a connected bead-spring chain, with $N_m$ uniformly charged monomers held together by a spring potential $U_b$, given as 
\begin{equation}
 U_{b} = \sum_{i=1}^{N_m-1}\frac{k_s}{2} \left( { r}_{i,i+1} -l_0 \right)^2,
\end{equation}
where $l_0$, $k_s$, ${\bf r}_i$, and  ${r}_{i,i+1} =
|{\bf r}_{i+1} - {\bf r}_{i}|$ denote the equilibrium bond length, the spring constant,  position of the $i^{th}$ monomer, and magnitude of  the bond vector connecting $i^{th}$ and ${(i+1)}^{th}$ monomers, respectively.

The excluded volume interaction among all the entities are incorporated via repulsive shifted  
Lennard-Jones (LJ) potential (often called the WCA potential) $U_{LJ}$ given as 
\begin{eqnarray}
 U_{LJ} =\sum_{i>j}^N  4 \epsilon_{LJ} \left[ \left(\frac{ \sigma}{r_{i,j}}\right)^{12} - 
 \left(\frac{\sigma}{r_{i,j}}\right)^{6} + \frac{1}{4} \right],
 \label{Eq:u_lj}
\end{eqnarray}
for $r_{i,j} \leq2^{1/6}\sigma$, 
and $U_{LJ}=0$, otherwise. 
 Here, $\sigma$ and $\epsilon_{LJ}$ are  the size of the bead and  interaction energy parameter, respectively. $N=N_m+N_m^c+N_s$ is the total number of particles that includes $N_m$ polymer monomers, $N_m^c$ counterions, and $N_s$ salt-ions  in the solution, all having same size $\sigma$. Similarly, the electrostatic interaction among all the charged entities is accounted using the long-range Coulomb 
 potential $U_c$, 
\begin{equation}
U_c= {\frac{1}{4\pi \varepsilon_0 \varepsilon_r} } 
\sum_{\mathbf{n}}{}^{'} \sum_{i=1}^{N} {\sum_{j=1}^{N}} \frac{q_i q_j}{|\mathbf{ r}_{i,j}+\mathbf{n}L|}. 
\end{equation} 
Additionally, the summation is performed over all the images  placed in periodic boxes positioned at $\mathbf{n}L$ in the 3D space, where L is the  length of  primary box. Symbol ' signifies the exclusion of self-interaction term  for the case  of $i=j$ if and only if $\mathbf{n}=0$. Here, $q_i$ and $q_j$ represents the charges which in case of the monomer is a negative unit charge $Z_m=-1$, counterions is  $Z_c=+1$ and for added salt cations and anions is $Z_s^+=+z$ and $Z_s^-=-1$ , respectively, where $z=1,2,3,4$ denotes the valency. The electrostatic strength is paramterized using the dimensionless number $\Gamma={l_B q_i q_j}/l_0$, where Bjerrum length, $l_B={\frac{1}{4\pi \varepsilon_0 \varepsilon_rk_BT} } $ is the length at which electrostatic energy is comparable to the thermal energy. In the physiological conditions for biological polymers like dsDNA, typically $\Gamma$ spans the range of  $2-4$ in the solution\cite{Muthu_Book_2011}. Here, we choose $\Gamma$ over a range $2-20$ throughout. In this range PE acquires conformations like stretched chain, coil and globule. 
However, most of the simulations are focused for $\Gamma=2-3$.
The Coulomb interactions are implemented using the standard 3D Ewald summation technique for the ionic mixtures in bulk\cite{deserno1998mesh_1}. For the simplicity of simulations dielectric constant of the medium, $\varepsilon = \varepsilon_0 \varepsilon_r$, is taken uniform throughout the polymer solution.  {\cblue However, it's possible to consider smaller $\varepsilon_r$ near the chain to mimic the realistic situation\cite{fahrenberger2014computing,fahrenberger2015influence,smiatek2020theoretical,tyagi2010iterative}. }
Solvent is modelled within the framework of the multi-particle collision dynamics (MPC)\cite{Kapral_ACP_2008,Gompper_APS_2009}. This is an explicit solvent based technique which takes into account both thermal fluctuations and long-range hydrodynamic interactions. The details of the method can be found in the references cited here \cite{Kapral_ACP_2008,Gompper_APS_2009,Ripoll04,Malevanets_MSM_1999,Lamura_EPL_2001,Singh_JPCM_2012}. 

All the physical parameters  are presented in basic system units of length $l_0$, energy $k_B T$, and time $\tau=\sqrt{m_s l_0^2/k_BT}$, where $m_s$ is mass of a solvent particle, i.e., unity. The MPC parameters are collision time $\tau_s=0.1 \tau$, cell length $a=l_0$ and average number of particles in a cell $\rho_s=10m_s/l_0^3$. This pertains to viscosity of the medium $\eta_s = 8.7 \sqrt{m_s k_B T/l^4}$. That gives the diffusivity of a free ion $D_0\approx 0.02$, and the associated time scale of diffusion $\tau_D=\sigma^2/D_0\approx 32$.  Parameters like spring constant  $k_s=10,000 k_BT/l_0^2 $, LJ energy $\epsilon_{LJ}/k_BT=1$, diameter of monomers, counterions and salt ions (all valencies) are taken to be $\sigma/l_0=0.8$, and their mass $M=10m_s$.  The chain length is chosen $N_m=50$ or $100$ in a cubic simulation box  of length $50$ and $64$, respectively.  The choice of  box length retains same monomer concentration $C_m\approx0.0004$. Additionally, the salt concentration is varied over a range of $C_s:0.00005-0.002$ for mono- and di-valent ions. 
Equations of motion of all the species are integrated using the  velocity-Verlet algorithm at a fixed integration time step $h_m=5\times10^{-3} \tau$.   

\begin{figure}[t]
\includegraphics[width=0.9\linewidth]{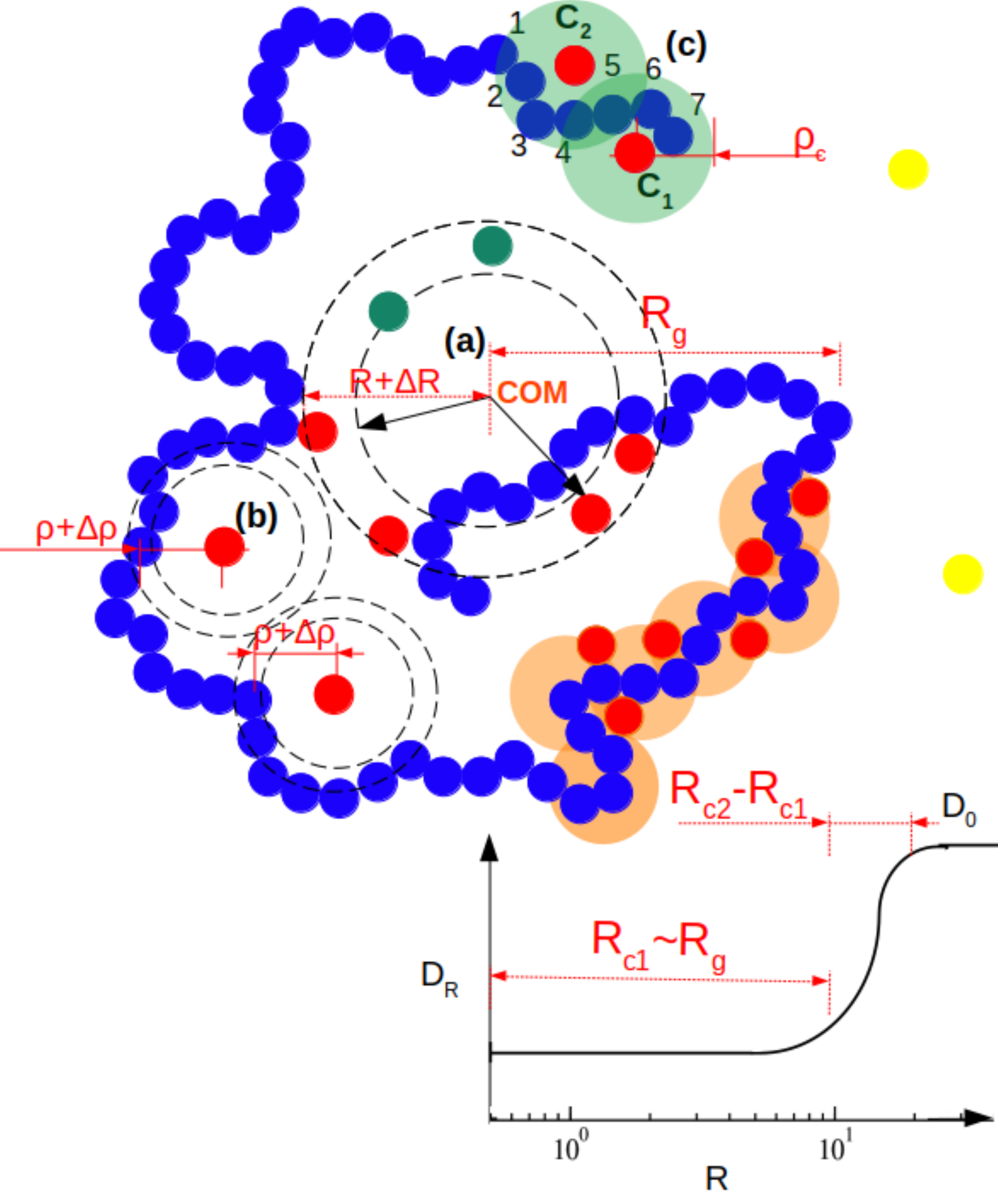}
\caption{An illustration of a PE chain and  characterization of ion-chain coupling at different scales; (a) w.r.t. COM: Ions within concentric shell $R$ to $R+\Delta R$ exercise effective diffusivity $D_R(R)$, (b) w.r.t. chain backbone: Ions situated at a distant $\rho$ to $\rho+\Delta\rho$ from their nearest chain segment possess the same diffusivity $D_{\rho}(\rho)$. Orange shaded region running along the chain demarcates the Stern layer consisting bound ions (red spheres) within $\rho<0.9$. Green spheres denote ions within chain boundaries, where $D_R<D_0$, while  yellow spheres represents free ions of diffusivity $D_0$. (c) {\textit{Effective charge per site:} Monomers are numbered 1-7 while counterions as $C_1$ and $C_2$. Charge of $C_1$ is distributed among monomers within $\rho_c<1.4$ i.e 4,5,6,7. Similarly, charge on $C_2$ is imparted to 1,2,3,4, and 5. Monomer indexed  5 possess a charge $z_5+\frac{z_{c1}}{4}+\frac{z_{c2}}{5}$, where $z_5$, $z_{c1}, z_{c2}$ represent charge on monomer 5, $C_1$ and $C_2$, respectively. } 
}\label{Fig:schematic}
\end{figure}

 \begin{figure}[t]
\includegraphics[width=\linewidth]{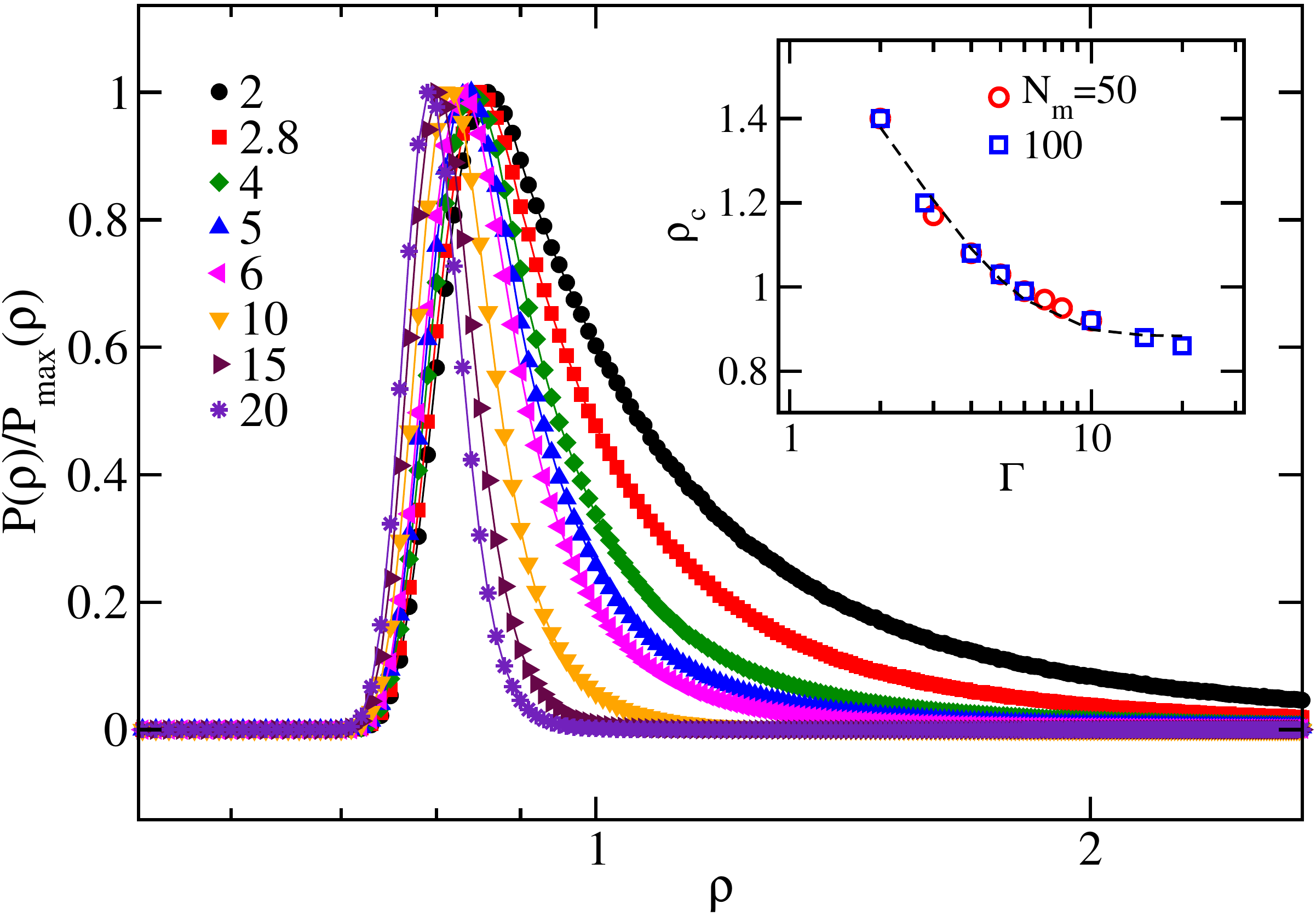}
\caption{ The scaled probability distribution $P(\rho)/P_{max}(\rho)$ of counterions as a function of its nearest distance from the chain $\rho$ for varying $\Gamma$ values, for chain length  $N_m=100$. $P_{max}(\rho)$ corresponding to the binding layer ranges from $\rho:0.9-0.8$ for $\Gamma:2-20$, respectively. Inset depicts the value of cutoff $\rho_c$ for $N_m=50$ and 100, where the probability density reduces to $80\%$ of its maximum value, demarcating the extend of the counterion cloud. }
\label{Fig:dist_ion}
\end{figure}




\section{ Distribution of Counterions}
In order to investigate various length-scales of ion-chain coupling, first we compute the density distribution $P(\rho)$ of the counterions as a function of its closest distance from the chain.  Here, $\rho$ is given as $\rho_i=  min(|\mathbf{r}_j-\mathbf{r}_i|) $, where $\mathbf{r}_i, \mathbf{r}_j$ corresponds to the position of the $i^{th}$ counterion and the $j^{th}$ monomer, respectively.  The distribution  shown in Fig.~\ref{Fig:dist_ion} provides a quantitative  estimate of the width of counterion cloud around the PE and demarcates the region of  adsorbed layer.  All the curves pertaining to  different  $\Gamma$  ranging from $2-20$ are  re-scaled by  corresponding $P_{max}$ value for the sake of comparative analysis.  The peak in distribution corresponds to the binding layer where ion-pair formations are prevalent. Beyond this length-scale, the effective screening from these ions weakens the interactive potential.  This results in a diminishing ion density profile into the solution. The strength of electrostatic attraction from the chain dictates the favourability of ion-pair formation and in turn the fraction of ions that are adsorbed  on the chain backbone. Further, Fig.\ref{Fig:dist_ion} (inset)   shows the cut-off distance ($\rho_c$) from the chain within which the probability has declined by $80\%$ of its maximum value. Evidently, the cutoff length decreases with increasing $\Gamma$, such that at $\Gamma=20$ with $\rho_c \sim 0.8$ almost all ions are bound to the chain. While for, $\Gamma=2$ where the electrostatic strength is very weak, the ion-cloud are loosely bound and extends beyond 
 $\rho_c \approx 1.4$. Additionally, the $\rho_c$ procured for $N_m=50$ and $N_m=100$ both superimpose on  each other. This indicates that the width of  the counterion cloud is primarily dictated by the strength of electrostatic interactions. 
 
 \begin{figure}
\includegraphics[width=\linewidth]{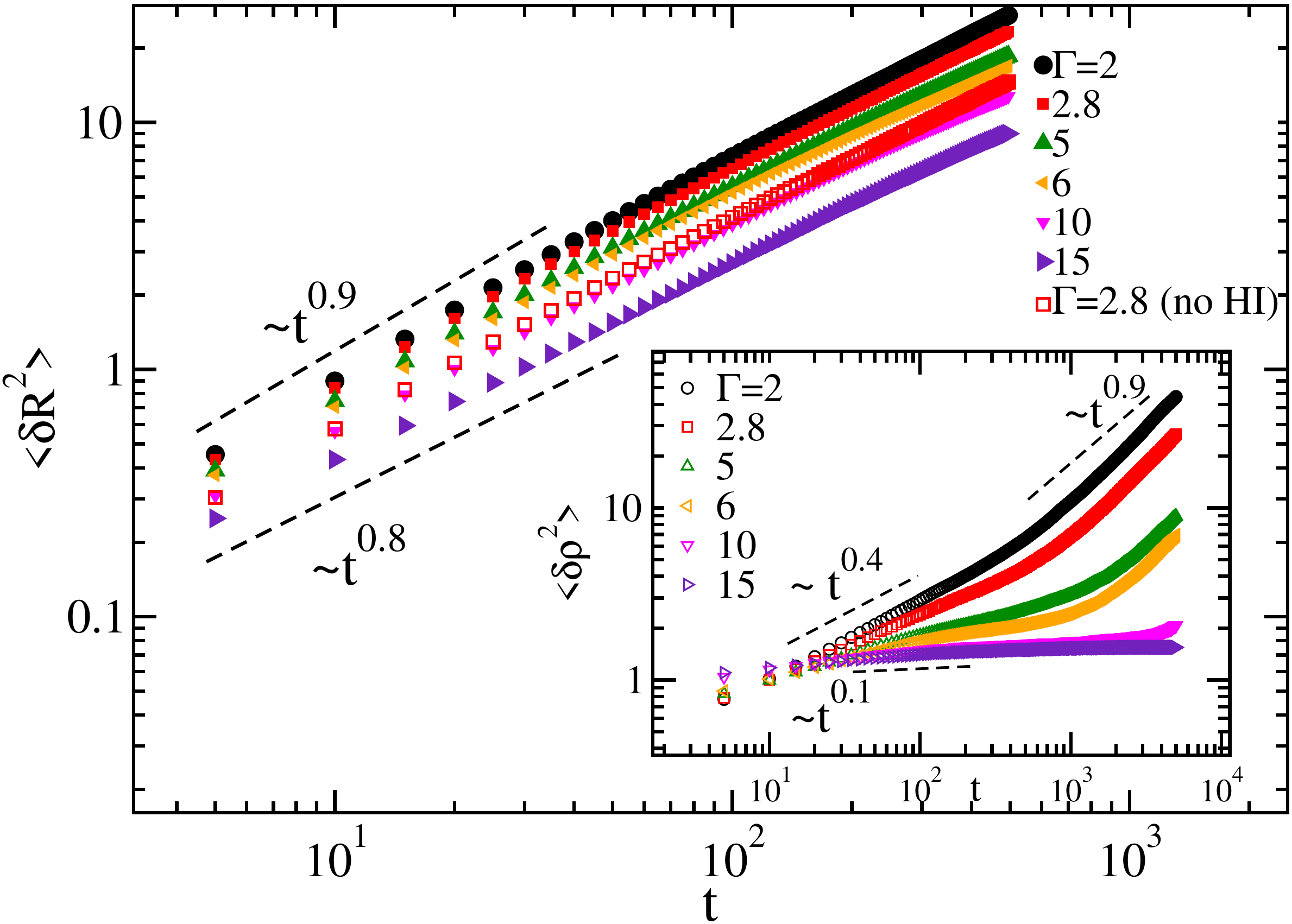}
\caption{The plot shows MSD of the adsorbed counterions w.r.t. chain's COM. The inset displays the MSD of ions in the transverse direction from the chain's backbone. The ensemble average is performed over those ions directly adhered to the chain backbone ($\rho<1.0$) at $t=0$. The dashed lines represent the sub-diffusive behavior observed in various regimes. {\cblue The open symbol (in red) in the main plot represents MSD for $\Gamma=2.8$, when  HI is absent. The later exhibits comparatively slower ion dynamics compared to the one in the presence of HI. } }
\label{Fig:msd_ions}
\end{figure}

\section{Transport Behavior of Counterions}


The electric double layer (EDL) around a macromolecule essentially constitutes of a Stern layer cohabited by the condensed ions, followed by a diffuse layer that falls off exponentially into the solution. 
Importantly, between the extreme limits of completely bound and free ions, an ion spends considerably large time traversing all across the chain neighbourhood pertaining to the interactive influence from the chain\cite{lifson1962self}. The emphasis here is to assess ion transport in terms of local diffusivity across the immediate vicinity of the chain's conformational span, and thereby, demarcate different coupling regimes associated with  the chain-counter ion duo. For that we compute, the  (MSD) and local diffusivity of adsorbed ions  relative to chain's COM and chain's segment in the following sections.


 \subsection{ Dynamics of Counterion }
  Looking at the ion dynamics from the chain's frame of reference is conducive in deciphering short scale local ion fluctuations, since ion coupling with chain's diffusion at larger timescales is necessarily suppressed. 
 In wake of this, first we estimate the relative diffusion of counterions with respect to chain's COM. This is computed from the MSD defined as,
\begin{equation}
<\delta R^2(t)> = <\frac{1}{N_a}\sum_{i=1}^{N_a}(\mathbf{R}_i(t)-\mathbf{R}_i(0))^2>,
\label{Eq:pair_diff_avg}
\end{equation}
where,
${\mathbf R}_i(t)= {\mathbf r}_i(t)-{\mathbf R}_{cm}(t) $ is the relative position of ion from the chain's COM at instant ${\text t}$. ${\mathbf r}_i(t)$, ${\mathbf R}_{cm}(t)$, $N_a$ are the position vector of $i^{th}$ counterion,  COM of the chain, and the number of adsorbed ions at $t=0$, respectively.  Angular bracket signifies average carried out over time and various ensembles. Here, the averaging is done over all the ions that are adsorbed on the chain at $t=0$. 

The estimated  MSD using Eq.~\ref{Eq:pair_diff_avg} is shown in Fig.~\ref{Fig:msd_ions}. Interestingly, the MSD curve exhibits sub-linearity with time as $<\delta R^2(t)>\sim t^{0.9}$,  even for time much longer than the diffusive time scale of a free-ion $\tau_D \approx 32$. The average MSD of ions for much longer times is shown in SI-Fig.1. A gradual crossover into a plateau regime is seen at large times, where the chain limits the movement of ions within 
it's proximity (see SI-Fig.1). This saturation in the ion dynamics happens roughly at  timescales when the chain has diffused its own size i.e, $t_p={R_g}^2/6D_p$. 
Hence, this indicates that beyond timescales of polymer diffusion the ion transport is strongly featured by the characteristics of  native PE dynamics, whereas at smaller timescales, the counterions exercise larger diffusional freedom. 

 Further, the self diffusion of ions within the chain's proximity  can be essentially resolved into two separate modes. One is ion traversing perpendicularly outwards from the chain backbone, while other is hopping/sliding motion of ion along the chain.   One way to distinctly obtain the transverse movement is by tracing an ion's diffusion w.r.t. the chain backbone itself instead of the COM, such that the diffusivity along the chain is suppressed. This  MSD  of ion from the chain backbone is calculated as, 
\begin{equation} \langle { \delta \rho}^2(t)\rangle ={\langle({\bf {{\boldsymbol  \rho}_i}}(t)-{{\boldsymbol  \rho}_i}(0))^2 \rangle}_{\rho} ,
\end{equation} 
here,  $i^{th}$ ion's closest distance from the chain is defined $\rho_i=  min(|\mathbf{r}_j-\mathbf{r}_i|) $, where $\mathbf{r}_i, \mathbf{r}_j$ corresponds to the position of the counterion and the chain monomer, respectively. The MSD characterization here is done as a function of $\rho$ with an ensemble average carried out for all ions present in the range of  $\rho(0)<1.0$ at $t=0$.  

This estimated MSD w.r.t. the chain's backbone $\langle \delta \rho^2(t)\rangle$ as prescribed above for various $\Gamma$ are shown in inset of Fig.~\ref{Fig:msd_ions}. 
A striking observation is that the transverse movement of these ions across the chain showcases a comparatively much slower diffusive mode, $\langle \delta \rho^2(t)\rangle\sim t^{0.4}$ relative to $\langle \delta R^2(t)\rangle\sim  t^{0.9}$ for $\Gamma=2.8$. This retardation in the ion dynamics outwards from the chain backbone further enhances for increasing $\Gamma$ strength as suggested by the plot, where for $\Gamma=10$, the MSD exhibits $t^{0.1}$ dependence. Once the ion traverses significant distance from chain's direct influence, a diffusive behavior   will be recovered at longer times. 
Despite, an overall effective diffusion ($\langle \delta R^2(t)\rangle\sim t^{0.9}$) exhibited by an ion near the chain, the suppression of transverse ion movement from the chain backbone indicates bound ion's strong diffusional proclivity along the chain contour.

{\cblue Further, the suppressed diffusivity seen in the transverse direction ($\perp$) due to the strong interactive influence of the chain, leads to large entrapment times of ions near the chain. This substantiates the occurrence of slow relaxation modes (called SLF modes in dielectric relaxation) of ions reported in the previous studies on dielectric response of PE  solution\cite{kim2015barrier,fischer2013low}. The SLF mode corresponds to the longest relaxation time associated with an ion that constitutes its transverse diffusion ($\perp$) from the chain, its residence time within the conformational boundary and its escape from the chain bound to hop onto nearby PE chains.    }
The enhanced residence time of ions on the chain backbone is addressed in detail in the  subsequent sections.  


 \subsection{Effective diffusivity relative to PE segment}
 In the previous section, we have estimated MSD of ions in the chain's periphery. Now, we translate these to quantify the effective transport coefficient of the ions. One way to look at the ion diffusivity in a  microscopic way is to characterize them based on their nearest distance from the chain. 
 This arises from the assumption that any two ions equally spaced from their respective nearest chain segments possesses the same diffusivity under same coupling strength. 
 
 \begin{figure}[t]
\includegraphics[width=\linewidth]{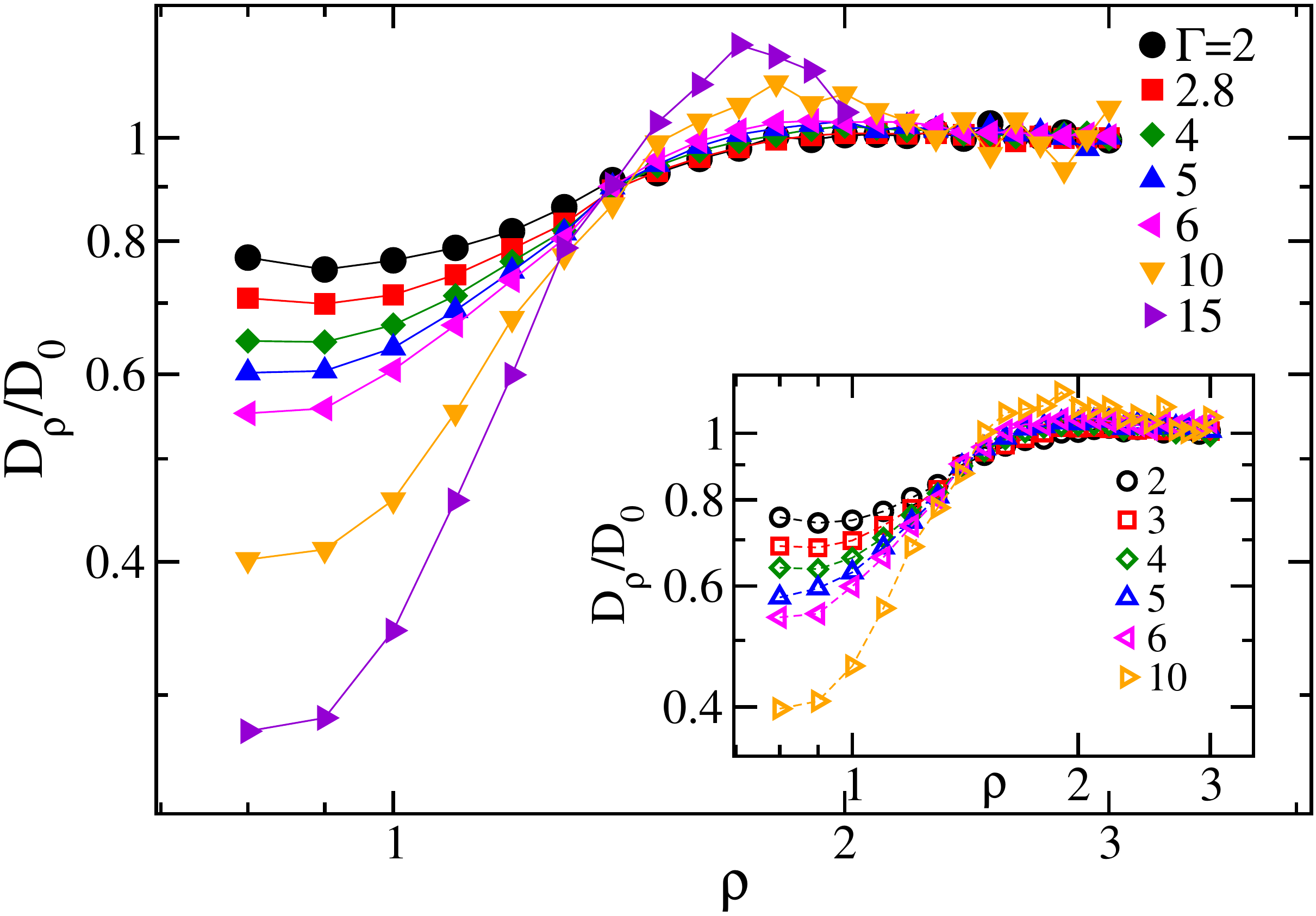}
\caption{ The effective diffusion coefficient $D_\rho/D_0$ of the counterion as a function of its nearest distance  $\rho$ from the chain backbone for polymer length $N_m=100$ (closed symbols) and inset shows for $N_m=50$ (open symbols). Here, $D_0\approx 0.02$ corresponds to the free ion diffusivity. Different curves pertain to varying values of $\Gamma$,  } 
\label{Fig:diff_tube}
\end{figure} For this,  we assume  concentric tubes of varying radius $\rho$ that runs along the chain segment. Such that the MSD of counterion w.r.t the chain's COM is  functionalized in terms of $\rho$, which follows from the expression, 
 
 \begin{equation}
<\delta R^2(t)> =  <\frac{1}{N_{\rho}}\sum_{i=1}^{N_{\rho}}(\mathbf{R}_i(t)-\mathbf{R}_i(0))^2>_{\rho}.
\label{Eq:pair_diff_com1}
\end{equation}

 Here, the ensemble average is done over all those  ions $N_{\rho}$ within the tube enclosing $\rho$ to $\rho+\Delta \rho$, where the thickness of concentric tubes is taken $\Delta \rho=0.1$, see Fig.\ref{Fig:schematic} for visualization. The effective diffusion coefficient  to characterize the ion-chain pair dynamics  at a given separation $\rho$ is procured by obtaining a fit 
$\langle  \delta R^2 \rangle \sim 6 D_{\rho} t$ to the MSD function. {\cblue  Mentionably, the effective diffusivity of the ions here entails the effect of the strong electrostatic attraction from the chain as well.} The fitting is performed in the region, where the diffused length of the pair does not exceed $<\delta R^2(t)>\approx 1$.   The normalized  local diffusivity  $D_\rho/D_0$ for various $\Gamma$ at  $N_m=100$  is plotted as a function of ion's nearest distance  $\rho$ in  Fig.~\ref{Fig:diff_tube}. The procured results reveal that for a given $\Gamma$,  the ion exercises a smooth and continuous increase in diffusivity, as they transition from being chain bound to completely free. The ion possesses the lowest diffusivity at a spatial separation of $\rho< 0.9$ from the chain backbone, which demarcates the \textit{Stern layer}, see Fig.\ref{Fig:schematic}. This is intuitive as any ion directly grazing the chain backbone is at  least at a separation $\sigma=0.8$. Beyond $\sigma$,  electrostatic screening caused by the ions residing on the chain, leads to a quick ascent in the effective diffusivity of ions. That further saturates  to free ions diffusivity beyond $\rho \sim 1.5$.  The chain's effective diffusion coefficient obtained for $N_m=50$ is compared with $N_m=100$ for a range of varying $\Gamma$  in inset of Fig.~\ref{Fig:diff_tube} showed in open symbols. Interestingly, the effective diffusivity $D_\rho/D_0$ is independent of the molecular weight for a given $\Gamma$ as depicted in Fig.~\ref{Fig:diff_tube}.   $D_\rho$ attains a peak in the narrow window of $\rho$ for larger $\Gamma=15$ and $20$, which might be due to the strong electrostatic attraction of ions onto the PE globular surface. 



 \begin{figure}[t]
\includegraphics[width=0.96\linewidth]{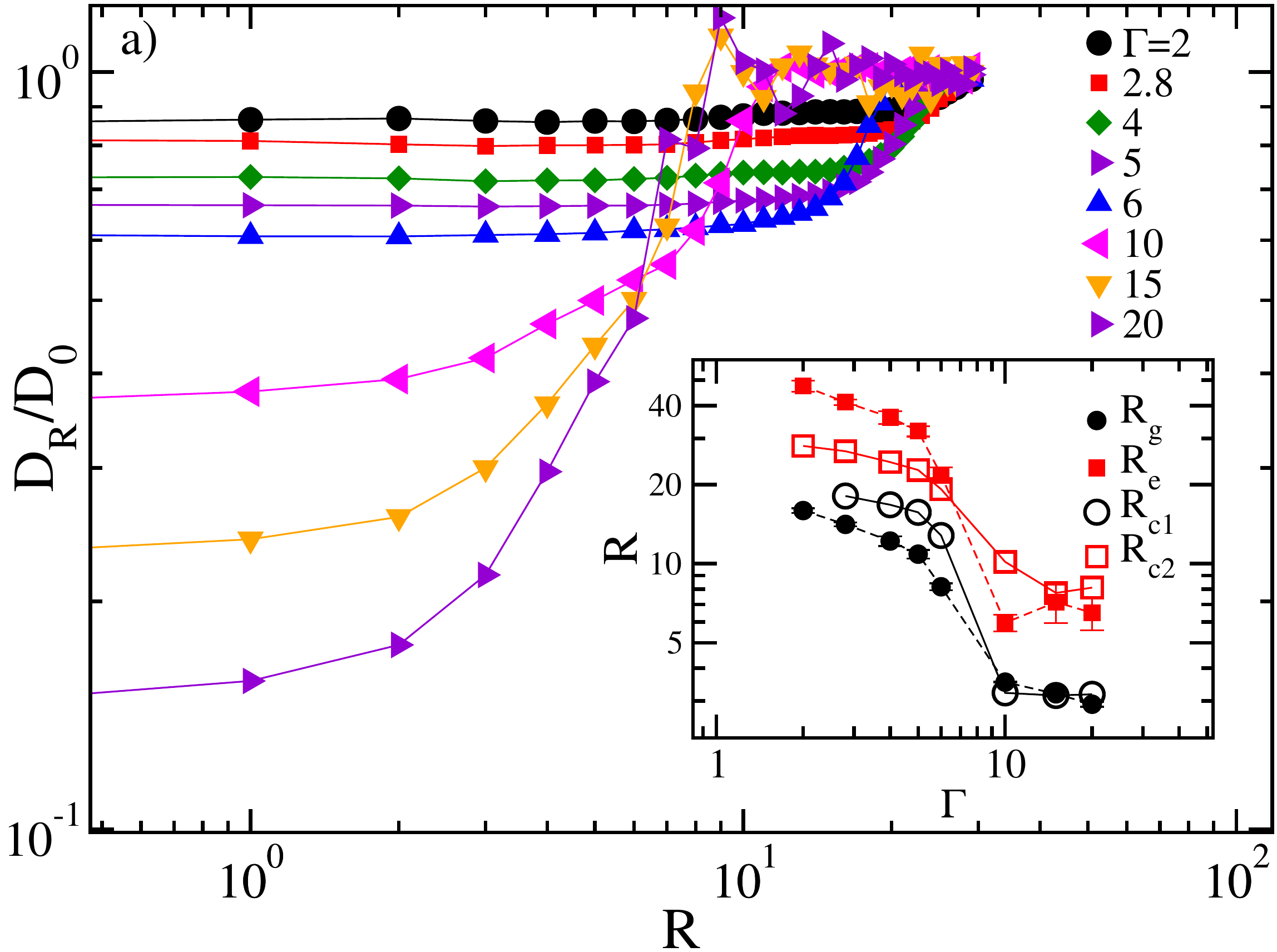}
\includegraphics[width=\linewidth]{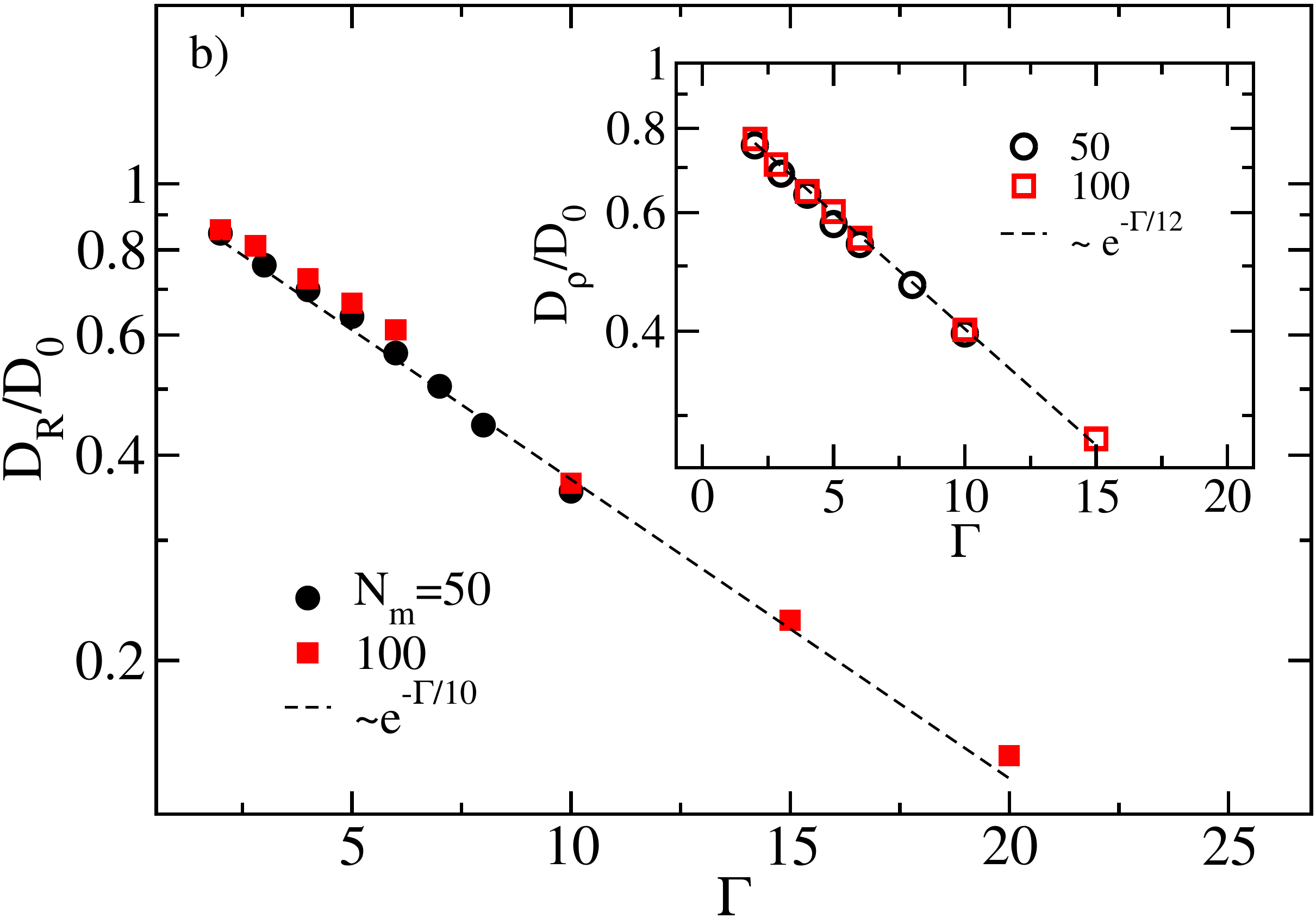}
\caption{ a) The effective  diffusion coefficient $D_R$ of the counterions w.r.t. COM as a function of the radial distance 
$ R= |r-R_{cm}|$ for $N_m=100$. Inset shows the $R_{c1}$, $R_{c2}$, $R_g$, and $R_e$ obtained from the main plot.
b) The effective  diffusion $D_R(0)$ of entraped ion within the chain's conformational boundary $R < R_g$  as a function of electrostatic strength $\Gamma$ for chain $N_m=50$ and 100. The dashed line shows the exponential fit $D_R \sim \exp(-\Gamma/a_R); a_R\approx 10$ to the data points. Inset: The normalised diffusivity of bound ions within the Stern layer ($\rho<0.9$) as a function of $\Gamma$ for varying chain lengths. The acquired exponential fit ($D_\rho \sim \exp(-\Gamma/a_\rho); a_\rho \approx 12$) is shown in dashed line.}
\label{Fig:pair_diff}
\end{figure}

 \subsection{Effective diffusion relative to chain's COM}
 \label{subsec:D_wrt_chain}
 
 
 The choice of ion's radial distance from the chain's COM to characterize diffusivity primarily stems from the assumption that the average potential of a flexible chain assumes a spherical symmetry, which diminishes radially outward. Since a counterion in the chain's vicinity moves under the combined influence of both thermal agitation and electrostatic field from the PE, this symmetry is reflected in their dynamics as well. 
 The counterion's effective diffusivity is procured from  it's MSD in the COM frame, using a similar protocol explained in the aforementioned section. Here, the diffusivity $D_R(R)$ is parametrized as a function of its distance $R$ from the chain's COM.
 The  associated MSD function  have semblance   to Eq.~\ref{Eq:pair_diff_avg},
 \begin{equation}
<\delta R^2(t)> = <\frac{1}{N_{R}}\sum_{i=1}^{N_{R}}(\mathbf{R_{i}}(t)-\mathbf{R_{i}}(0))^2>_{R},
\label{Eq:pair_diff_com}
\end{equation}
with the only exception that the ensemble average is  over $N_R$ ions  starting off between $R$ to $R+\Delta R$ at initial time $t=0$. The $R$ space is discretised into concentric spherical shells centered at COM, where the shell thickness is taken $\Delta R=1.0$, see Fig.\ref{Fig:schematic} for visualization. 
The procured values of $D_R$ normalised w.r.t. $D_0$ as a function of radial separation $R$  is presented in Fig.~\ref{Fig:pair_diff}-a for various $\Gamma$ at  $N_m=100$ (see SI-Fig.2 for $N_m=50$).  
For each of the $\Gamma$ curves, up to a certain distance from the COM the $D_R$ remains nearly constant constituting a plateau region, then followed by a transition regime where the diffusivity gradually grows before hitting a plateau of free ions diffusivity.

Here, we demarcate the transition in diffusivity at a distance $R_{c1}$, where $D_R$ becomes $D_R=D_R(0)+0.1*(D_0-D_R(0)))$, see schematic in Fig.\ref{Fig:schematic} for the definition of $R_{c1}$. Interestingly, the radial cutoff $R_{c1}$, at which the plateau region pertaining to the lowest diffusivity of adsorbed ions deviates from its constant value, is comparable to the radius of gyration $R_g$ as shown in the inset of Fig.~\ref{Fig:pair_diff}-a. Here, $R_g$ stands for the radius of gyration of the chain given as, $R_g^2={\langle \frac{1}{N_m}\sum_{i=1}^{N_m}( {\mathbf R_i}-{\mathbf R}_{cm})^2\rangle}$. 
This coupling of ion with the chain at the macroscopic length scale can be vividly explained considering how  an ion traverses within the chain's proximity. 
For example at one instance, an ion positioned at $R$ might have a sufficiently large interspatial separation from the chain backbone; however, with time it might traverse to a position where it is still nearly at $R$, but comes in direct contact with a chain segment (see schematic in Fig.\ref{Fig:schematic}). Hence, ions in the vicinity of a flexible polymer are entrapped within the conformational boundaries for longer times\cite{lifson1962self}, due to this frequent contacts with the fluctuating chain segments. Within $ R_{c1} \sim R_g$ where  fluctuating polymer coil encompasses a spherically symmetric volume in terms of the monomer density distribution, an ion executes the lowest diffusive mode homogeneous across the region.  This regime hence can be termed as the core of monomer-counterion coupling.


Further beyond $ R_{c1}$, where diffusivity gradually ascents, we assume a cut-off $ R_{c2}$, where the value of $D_R=D_R(0)+0.8*(D_0-D_R(0)))$ such that $R_{c2}-R_{c1}$ parametrizes the width of this transition regime. This $R_{c2}$ is comparable to another important length scale of the system, i.e., mean end-to-end distance $R_e$ given as $R_e=\sqrt{\langle ( {\mathbf R}_1-{\mathbf R}_N)^2\rangle}$, as shown in the inset of Fig.~\ref{Fig:pair_diff}-a. Although at lower $\Gamma$'s, the association of diffusivity to chain expanse is weak ($R_{c2}<R_e$), as the chain attains a comparatively open structure leading to less segmental fluctuations. This suggests that beyond the conformational span of the chain $ R_{c1}\sim R_g$, the ion-distribution is dictated by a screening potential with a characteristic width $ R_{c2}- R_{c1}$.  
The $ R_{c2} - R_{c1}$  simply demarcates the width of \textit{diffuse layer} around a flexible chain, where the ion diffusivity enhances radially outwards following a weakening of the ion-chain coupling into the solution.

In the preceding case of ion characterization relative to the chain segment, diffusivity $D_\rho$ of an ion is independent of molecular weight over the whole range of $\rho$. However, here $D_R$  exhibits dependence on the chain  macroscopic length scales $R_e$ and $R_g$, see inset of Fig.~\ref{Fig:pair_diff}-a.  
This is because  characterizing diffusivity using tube symmetry,  the association of ions with chain segmental fluctuation is obscured due to tube running along the chain. As a result,  retardation of diffusivity occurs at smaller scales. Further, Fig.~\ref{Fig:pair_diff}-b displays the effective diffusivity $D_R(0)/D_0$ of entraped ions within chain conformational boundary $R<R_g$, which  exhibits an exponential dependence on $\Gamma$ and is independent of the molecular weight. Similar, values are procured for the diffusivity of ions directly bound to the chain backbone within $\rho<1$ (see inset of Fig.~\ref{Fig:pair_diff}-b ), since ions entraped within the chain proximity in effect exercise diffusion close to that of bound ions due to frequent adsorptions on the chain segment.    








\section{Binding time}
\label{sec:binding_time}
While the preceding section mainly deciphers the relevant length scales of ion-chain complex based on ion transport properties across these regimes, this section emphasizes on the ion's association time with the chain. 
In order to get a closer insight regarding their spatial evolution w.r.t. the chain backbone, we tag a counterion and trace its nearest distance from the chain backbone over progressing time. This is elucidated in SI-Fig.-3, sec. I.  

In wake of the nature of fluctuations seen, we can primarily classify three distinct ionic events; (a) {\bf Ion adsorption} : within $\rho_{c1}$, when an ion is directly bound to the chain, (b) {\bf Ion desorption}: within $\rho_{c1}<\rho<\rho_{c2}$, when it detaches from the chain backbone, but yet remains in chain's coupling proximity, where $D_\rho<D_0$ (see Fig.~\ref{Fig:diff_tube}), and (c) {\bf Free ion: } when ion navigates beyond $\rho_{c2}$, where $D_ \rho \approx  D_0$, it decouples from the chain exercising translational freedom. The time elapsed between a counterion entering first time in $\rho<\rho_{c1}$ and leaving the same is parametrized as the ion binding time $t_b$\cite{shi2021counterion,cui2007counterion,karatasos2009dynamics}. Additionally,  $\rho_{c1}$  includes the fluctuation of ion involved in the bound state and also must not exceed beyond a certain $\rho$, where the desorption and the adsorption events are not discernible. Considering that for $\Gamma:2-20$ the binding layer spans the range $0.8-0.9$ and width of ion-cloud in the chain proximity ranges from $0.9-1.4$ (shown in Fig.~\ref{Fig:dist_ion}-b, inset), a reasonable choice for $\rho_{c1}$ can be anywhere between $\rho_{c1}= 0.8 - 1.4$.

\begin{figure}[t]
\includegraphics[width=\linewidth]{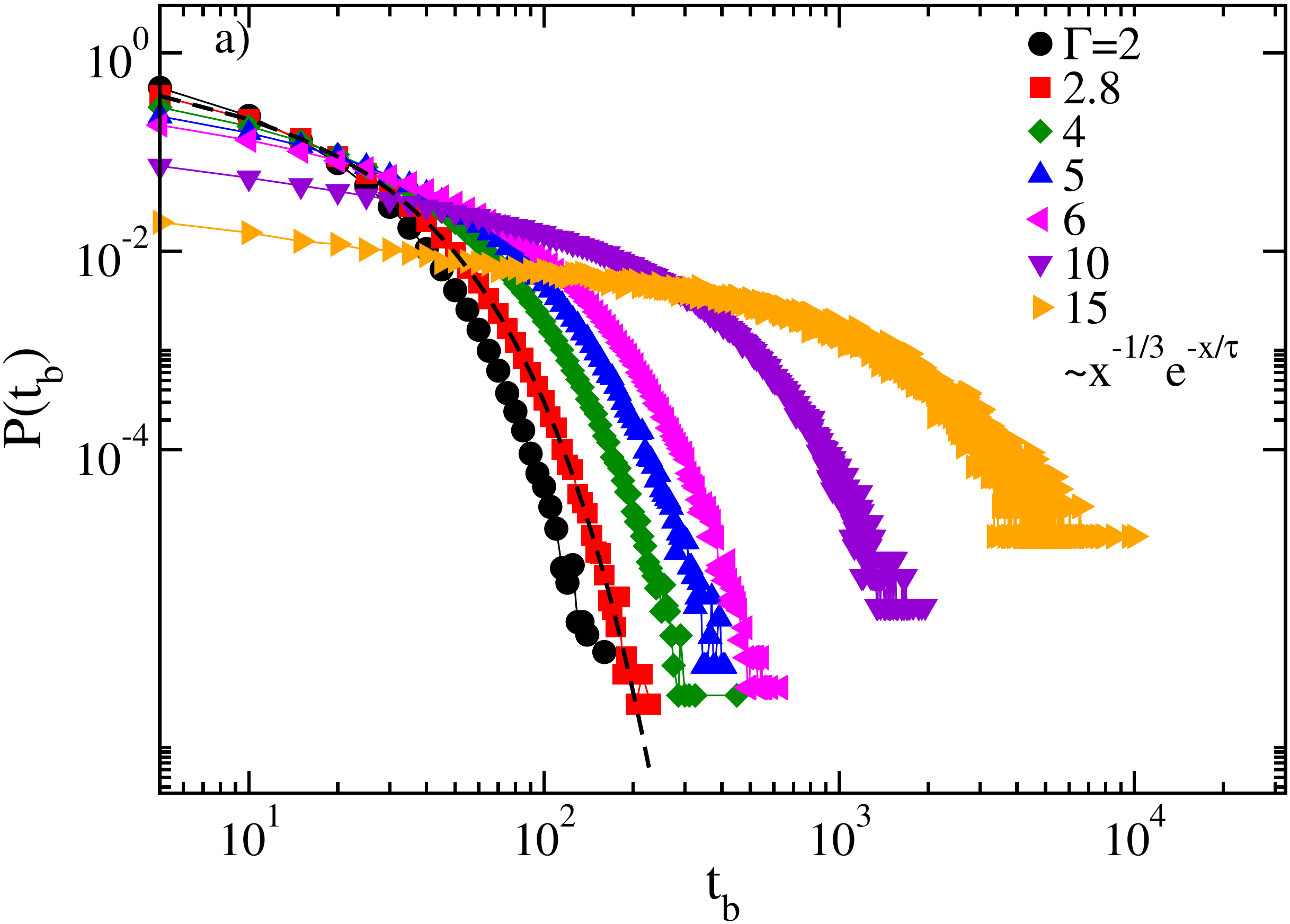}
\includegraphics[width=\linewidth]{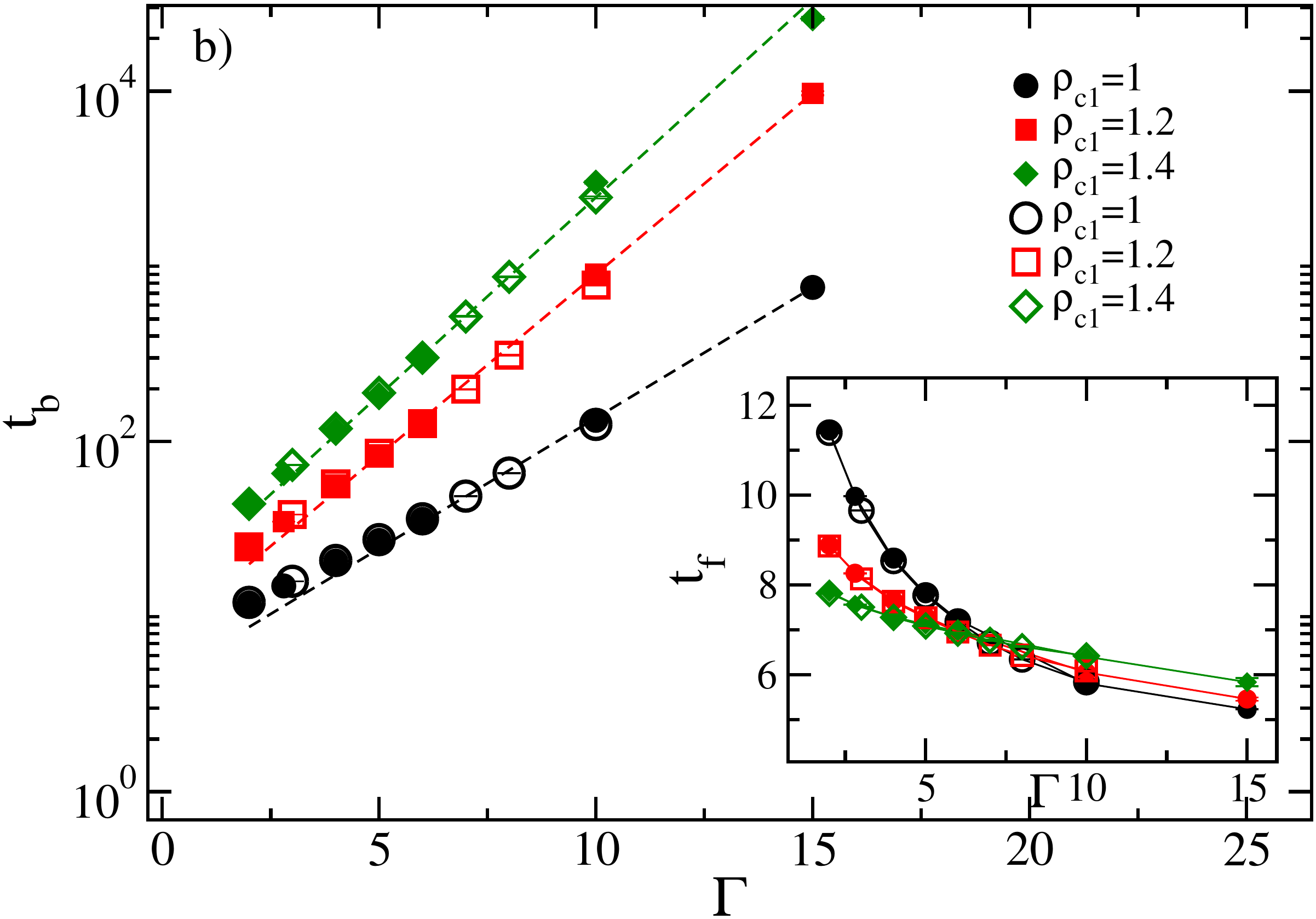}

\caption{ a) The probability distribution function  $P(t_b)$ of  binding time within cutoff $\rho_c=1.0$ procured for varying $\Gamma$, represented in different colors for chain length $N_m=100$. A fit  obtained using expression $t^{-1/3}\exp(-t/\tau_b)$, is shown by dashed line for $\Gamma=8$. 
b) Binding time $t_b$ with increasing $\Gamma$ for few chosen cutoff-distances $\rho_{c1}=1.0$, 1.2, and 1.4 for $N_m=100$ chain length (closed symbol) and $N_m=50$ (open symbol). At lower $\Gamma<3$, $t_b$ is comparable to the ion diffusive scale $\sigma^2/D_0\approx 32$. {\cblue The dashed lines show the exponential fit to the data.} Inset shows ion's free time $t_f$ within $\rho_{c1}<\rho<\rho_{c2}=2.0$ at $\rho_{c1}=1.0,1.2,$ and $1.4$ for $N_m=100$ (closed symbol) and $N_m=50$ (open symbol).  } 
\label{Fig:binding_time}
\end{figure}

 Figure~\ref{Fig:binding_time}-a shows the probability distribution $P(t_b)$ of binding time within a cutoff $\rho_{c1}=1$. Interestingly, the distribution yields two different regimes  with a power law decline in short time followed by an exponential decrease. The obtained fitting using function $P(t_b)\sim {t_b}^{-1/3}\exp(-{t_b}/\tau_b)$ for $\Gamma=2.8$  is represented by a dashed line. A comparatively larger probability at shorter binding times, reflected as $t^{-1/3}$ dependence, is an inadvertent consequence of the fluctuation of ion's situated near the cutoff\cite{cui2007counterion}.  
 {\cblue A power law behavior of the counterion residence time is also reported in the atomistic simulations with relatively larger exponent\cite{heyda2012ion}.  }
 The average binding time $t_b$ estimated from this distribution as a function of $\Gamma$ is shown in Fig.~\ref{Fig:binding_time}-b  at different $\rho_{c1}$ values for chain length $N_m=100$ (shown in closed symbols). Interestingly, the binding time exhibits an exponential increase with  $\Gamma $. This stems from ions own diffusivity within the bound layer, which decreases exponentially with $\Gamma$, as a result of stronger coupling (see Fig.\ref{Fig:pair_diff}-b). This smaller diffusivity of adsorbed ions translates as the enhanced residence time of ions near the chain. For larger binding cutoffs ($\rho_{c1}$), the ion association time increases as short-time ionic fluctuations get suppressed. Additionally, $t_b$ values procured for chain length $N_m=50$ (shown in open symbols) are equivalent to $N_m=100$ case, for all cutoffs.   This we speculate  is a consequence of the same ionic concentration found in both chain lengths $N_m=50$  and $100$, since the exchange rate ($\sim 1/t_b$) of ions across the Stern layer primarily depends on both the ion-monomer coupling strength and the local ionic concentration\cite{shi2021counterion}. The dependence of this microscopic ionic fluctuations on concentration is addressed in later sections.  Additionally, due to the constant exchange of ions across the binding layer, there is a definitive span an ion after desorption remains free before getting re-adsorbed again. For this free time estimation, only desorption events extending up to $\rho_{c1}<\rho<\rho_{c2}$ is accounted, where $\rho_{c2}\approx 2$. 

Beyond $\rho_{c2}$, an ion exercises nearly free diffusion and might escape conformational boundaries of the PE, hence it barely contributes to the exchange  across the binding layer. The estimated free time $t_f$ of ions is shown in Fig.~\ref{Fig:binding_time}-b (inset) for $N_m=100$ in closed symbols for varying $\rho_{c1}$ at  $\rho_{c2}=2$. Here, it accounts for the average time an ion navigates in $\rho_{c1}<\rho<\rho_{c2}$ from the instant it exits binding region $\rho_{c1}$, till it renters the same. For lower $\Gamma$, the binding time and the free time are comparable. While with increasing $\Gamma$  free time exhibits meagre variation unlike the binding time that grows exponentially. 
With increasing $\Gamma$, the larger electrostatic interaction  strength resulting in enrichment of ions on the chain backbone causes enhanced screening  beyond $\rho_{c1} > 1$. This screening with interactive strength is responsible for the minimal variation seen in $t_f$ of ions over increasing $\Gamma$. Free time estimated for $N_m=50$ is shown in open symbols, which again elucidates the same chain length independence. 

\section{The fluctuation of effective charge} 
The  previous sections have been mostly centered around the nature of ion movement under the interactive influence of the chain and the associated timescales. The current section  discusses how these ion fluctuations effectuate local charge regulation along the chain and its interesting consequences. 

\subsection{Effective charge }
In most of the previous studies\cite{Frank_EPL_2008,Liu_JCP_2002,Winkler_PRL_1998,Singh_2014_JCP,radhakrishnan2019force}, where the emphasis is on overall chain's degree of ionization, the total ionization state of the chain is evaluated as  $\alpha=(N_m-N_c)/N_m$. Here, $N_m$ and $N_c$ are total and adsorbed number of counterions, respectively, where a counterion is considered adsorbed below a certain cutoff $\rho_c$. This then imparts a uniform ionization to all repeat units along the chain, obliterating fluctuations in the ionization states of these binding sites. However,  to emphasize on the local ionization state and it's fluctuation,  the charge regularisation in the monomer-counterion vicinity needs to be considered. For that, we consider the charge effectuated by a counterion  is equally distributed among all the cohabiting monomers within a shell of $\rho <\rho_c$, with the bounding sphere centered at the reference counterion $j$. This is represented in the schematic shown in Fig.~\ref{Fig:schematic}, where the effective charge assigned to  $i^{th}$ monomer  is $Q_i=z_i+\sum_{j}{\frac{z_j}{N_j}}$.
Here, $z_i$ and $z_j$ denotes the absolute charge of $i^{th}$ monomer  and $j^{th}$ counterion, respectively. While summation runs over all the counterions indexed $j$ within the bounding sphere around monomer $i$, within it $N_j$ in turn denotes the number of cohabiting monomers within $\rho< \rho_c$ bound of counterion j. This effective charge estimation is pictorially represented in schematic shown  in Fig.\ref{Fig:schematic} c part. For example, monomer indexed  5 possess a charge $Q_5=z_5+\frac{z_{c1}}{N_{C1}}+\frac{z_{c2}}{N_{C2}}$, where counterion $C_1$ is shared among 4 monomers while $C_2$ is shared by 5 that gives  $N_{C1}=4$ and  $N_{C2}=5$. Here, we choose $\rho_c=1.4$ for the calculations of the effective charge across all $\Gamma$.


 \subsection{Effective charge: Spatial correlation}\label{sec::space_corr} 
 
 The local charge fluctuations along the chain backbone can be paramterized using the correlation of the effective charge per site as a function of its spatial separation.  The  correlation function  $C_{QQ}(\Delta r)$ is computed as,
 \begin{equation}
C_{QQ}(\Delta r)= \frac{<\delta Q(r_i+\Delta r)\delta Q(r_i)>	}
{<\delta Q(r_i)\delta Q(r_i)> }.
\end{equation}
Here, $r_i$ denotes position of the $i^{th}$ monomer site, $\Delta r=|\Vec{r_j}-\Vec{r_i}|$ denotes the inter-spatial separation between two such monomeric sites, and
$\delta Q(r_i) = Q(r_i)-\langle Q(r_i)\rangle$ corresponds to the fluctuation of effective charge at the monomer site positioned at $r_i$. The estimation of $Q(r_i)$ follows from the preceding section.

\begin{figure}
\includegraphics*[width=\linewidth]{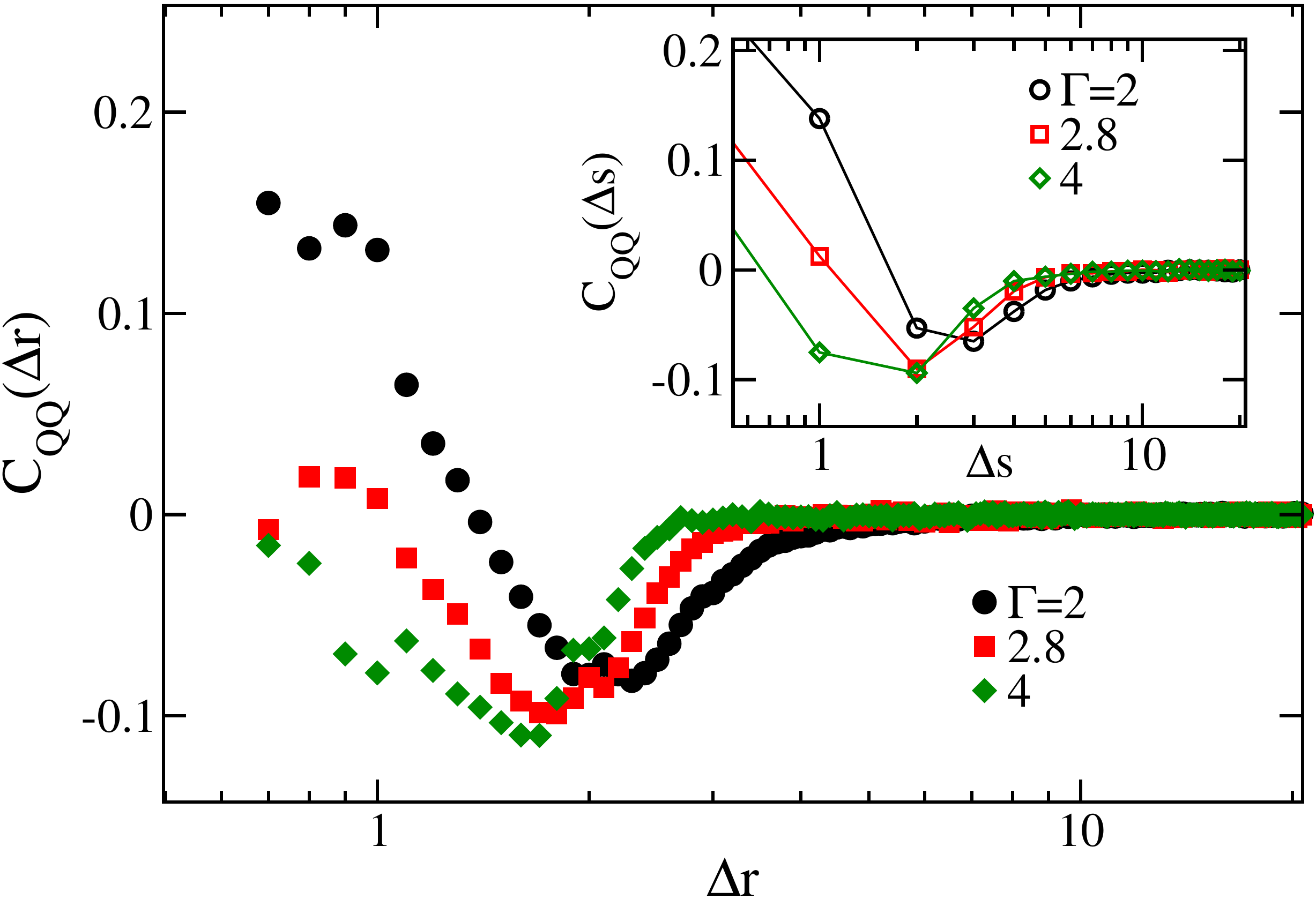}
\caption{Spatial charge correlation $C_{QQ}(\Delta r)$ as a function of inter site separation $\Delta r$ in the salt free  case for varying $\Gamma$. Inset pertains to the charge correlation $C_{QQ}(\Delta s)$ between monomeric sites estimated along the chain contour $\Delta s$, displaying similar behavior as in the main plot. }
\label{Fig:space_corr}
\end{figure}

 
 Figure~\ref{Fig:space_corr} displays  estimated effective spatial charge correlation as a function of $\Delta r $  for varying $\Gamma$. With increasing separation the correlation exhibits a quick decrease, but surprisingly beyond a certain $\Delta r$ instead of diminishing off to zero the correlation becomes negative. The strength of this negative correlation quickly relaxes, thus becoming zero again  beyond  $\Delta r \approx 4$, for  $\Gamma=2$. For larger $\Gamma$, the magnitude of negative correlation enhances, however the extent of $\Delta r$ over which it prevails shrinks, like for $\Gamma=4$,  $\Delta r$ reduces to 2.5. 
 
 The characteristic  features of $C_{QQ}$  can be understood by having a closer look on the charge fluctuation $\delta Q(r_i) = Q(r_i)-\langle Q(r_i)\rangle$, where $Q(r_i)$ falls in the range  $-1$ to $0$ for the monovalent ions. Here, $Q(r_i)=0$ and $-1$, signifies the neutralized and  completely ionized case, respectively. Therefore, at any given point of time if the instantaneous value of $Q(r_i)$ is such that $Q(r_i)>\langle Q(r_i)\rangle $ giving $\delta Q(r_i)>0$; it indicates an excess charge and if opposite prevails $\delta Q(r_i)<0$; indicates the local charge deficit.  Therefore, this negative correlation is a possible repercussion of the concomitance of an ion rich and ion deficit regions at certain interspatial separations. This local
fluctuation in ionic distribution might effectuate short-range attractive interactions, as reported in many multi-body complexation phenomenons\cite{angelini2005structure, kirkwood1952forces,blanco2019role,da2009polyelectrolyte,lund2013charge}.

\begin{figure}[thb!]
\includegraphics*[width=\linewidth]{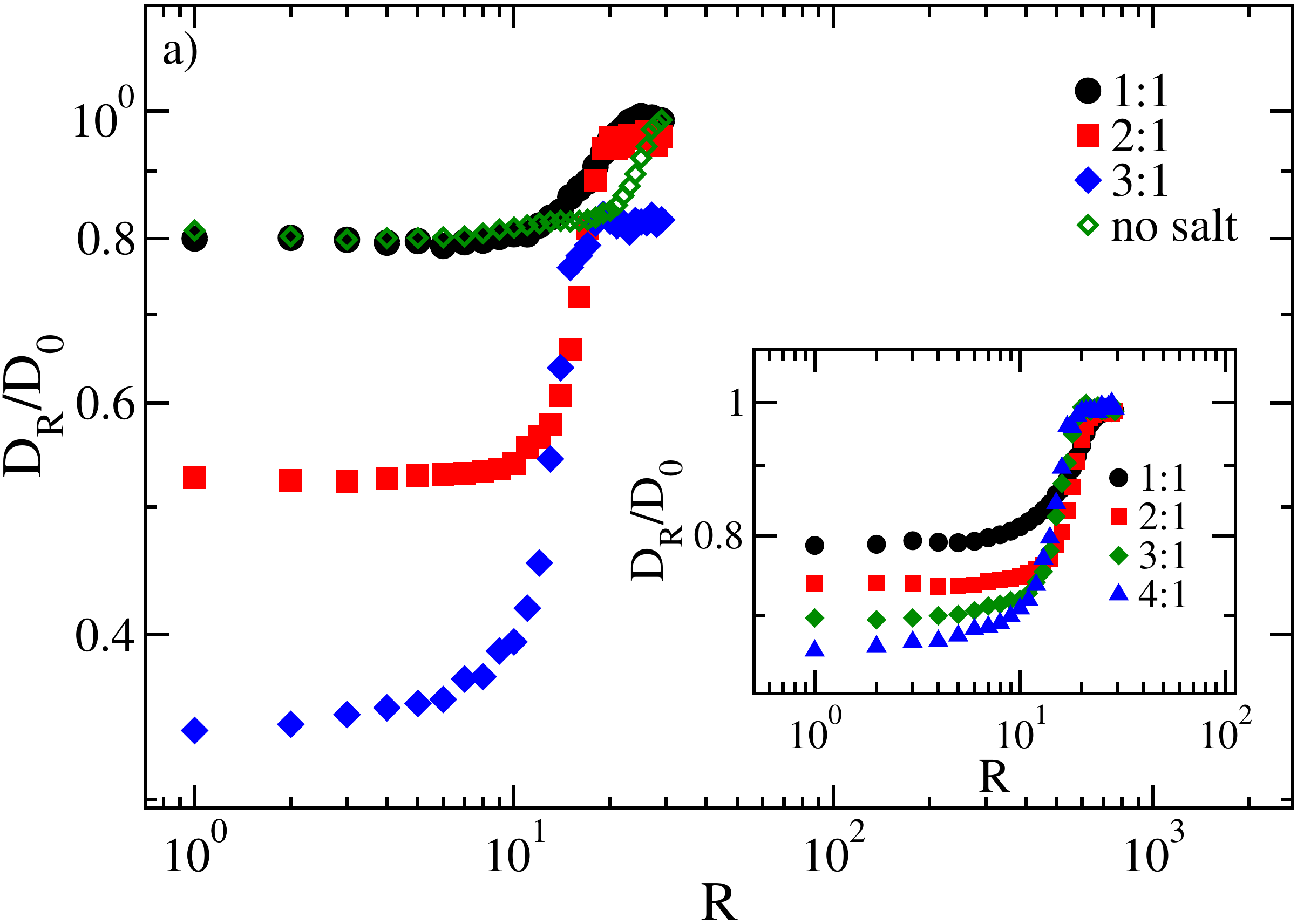}
\includegraphics*[width=\linewidth]{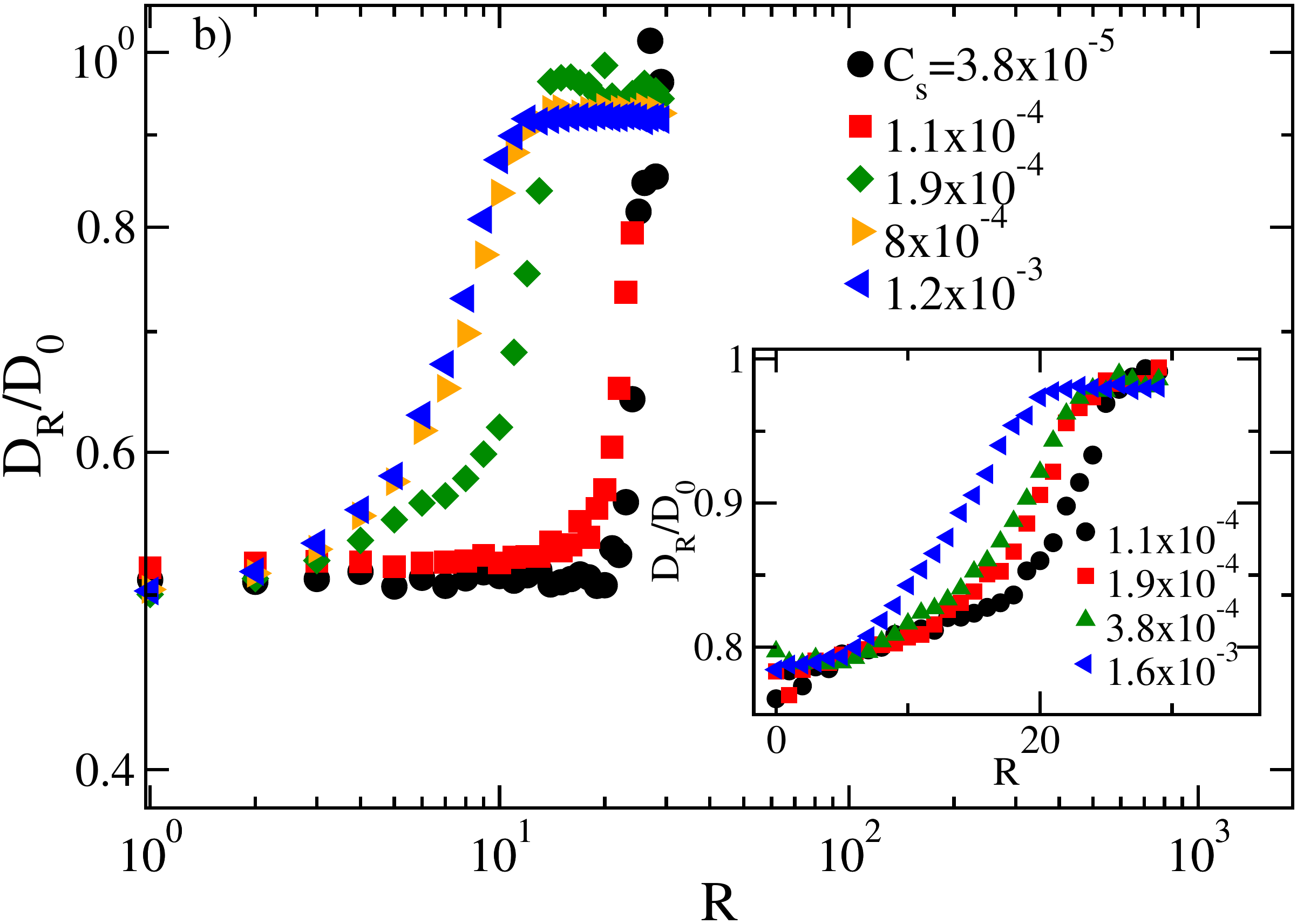}
\caption{a) The normalised effective diffusivity $D_R/D_0$ of added salt cations as a function of distance from the COM for $N_m=100$ at $\Gamma=3$. Individual curves pertains to different valencies of salt ion monovalent (1:1), divalent (2:1), and trivalent (3:1) for a fixed value of $\Gamma=3$ at $C_s=0.0004,0.0002$, and $0.0001$, respectively. The choice of salt concentration is such that the added salt have  same overall charge as the PE. Inset shows $D_R/D_0$ of salt cation as a function $R$  for varying valencies of added salt for a fixed value of interaction strength $\Gamma=3$. b) The effect of salt concentration on ion diffusivity $D_R/D_0$ for di-valent  salt and mono-valent  salt (inset) for $l_B=3.0$, corresponding to $\Gamma=3.0$ and $\Gamma=6.0$ for 1:1 and 2:1 salt cases, respectively . }
\label{fig:salt_effect}
\end{figure}

\section{Ion dynamics under added salt}
Presence of salt especially of higher valency, brings further intricacies  to the existing landscape of polymer-counterion complex. Like in view of electrostatic coupling strength and subsequent condensation on chain, the preponderance of higher valent ions over monovalent leads to a dramatic acceleration of the charge neutralization process. Such that size compaction can be easily achieved even at modest temperatures ( since $\Gamma$ increases $z_i$ folds), unlike the monovalent salt\cite{raspaud1998precipitation,sabbagh2000solubility, wittmer1995precipitation, kundagrami2008theory}. Several theories \cite{wittmer1995precipitation,kundagrami2008theory}, simulations\cite{hsiao2006salt} and experiments\cite{sabbagh2000solubility,raspaud1998precipitation,pelta1996dna,delsanti1994phase} on NaPSS and ssDNA have reported a re-entrant phenomenon, where the overcharging  due to higher valent salts ($z_i>1$) at moderate concentrations even leads to re-dissolution of the chain precipitate or re-expansion of a chain\cite{murayama2003elastic,hsiao2006salt}. Even the coions exhibit profound coupling within the ion atmosphere, leading to formation of ionic multilayers\cite{hsiao2006salt}.
Apart from the static properties, higher valency exhibits enhanced dynamic coupling with poly-ions, where diffusion of a PE is enhanced nearly twice in presence  of divalent salts\cite{chang2003brownian}.


 In purview of this, here we extend our study  of ion dynamics to the added salt scenario. Figure~\ref{fig:salt_effect}-a presents the effective diffusivity of the cations ($Z_s^{+z}$) of z:1 salt, as a function of ion's spatial separation from the chain's COM for a fixed $\Gamma=3.0$.  
 For the monovalent salt (1:1), the diffusivity of $Z_s^{+1}$ ions overlaps with the diffusivity of  counterions $Z_c^{+1}$ obtained in the absence of salt. This arises from the indistinguishability of counterions and salt cations. For divalent salt (2:1), the diffusivity of  cations  ($Z_s^{+2}$) exhibit a significant reduction, while the monovalent counterions still posses  diffusivity as in the case  of no salt. This is because for any higher valent salt the monomer-$Z_s^{+z}$ cation pairing is dictated by a coupling strength $\Gamma=\frac{zl_B}{l_0}$, which is z times larger than the  monomer-$Z_c^{+1}$ counterion pairing strength. The exponential suppression of bound ion's diffusivity with coupling strength is discussed in the salt free case~\ref{subsec:D_wrt_chain}. 
 
 Further, inset of Fig.~\ref{fig:salt_effect}-a  shows the variation in cation's diffusivity for the case of mono, di, tri, and tetra valent salt at fixed $\Gamma=3$ instead of fixed $l_B$. This arises from the consideration that, although with varying $l_B$ the solvent quality might be different, but the ion pairing strength dictated by $\Gamma$ in all salt solutions will be same. Interestingly, despite the same coupling  ion diffusivity exhibits a decrease (though less pronounced) with increasing valency. This can be attributed to the phenomenon called \textit{ion-bridging} reported in past studies\cite{kundagrami2008theory,huber1993calcium,schweins2001collapse}, where a divalent ion pairs up with two non-bonded monomeric sites, forming a bridge between them. This trans site multi-pairing of ion effectuates long range attractions within the polyion, and induces pronounced chain compaction even at modest temperatures. Similarly, with increasing valency of bridging cation, the networking happens between more monomers. Thereby, this enhanced electrostatic coupling via ion-bridging leads to further diffusional retardation at higher valencies.      

Besides valency, ion concentration  is another important metric that immensely influence the chain properties on a macroscopic scale\cite{hsiao2006salt}. Figure~\ref{fig:salt_effect}-b shows the spatial variation in diffusivity (relative to chain's  COM see section~\ref{subsec:D_wrt_chain}) of cations in divalent salt for a range of salt concentrations $C_s$. Surprisingly, the diffusivity of condensed ions is independent of salt concentration at short length scale, even though the diffusivity at higher spatial separation shows variation. This is a consequence of the chain acquiring open state to  globular state with increasing  salt concentration.   
The coupling of ion dynamics with the macroscopic boundaries reflects as different transition points for the diffusivity curve, where within it the ion transport remains unaltered to concentration variations. Similar behaviour is seen for the monovalent salt as shown in inset of Fig.~\ref{fig:salt_effect}-b.  We have compared the length-scale corresponding to the ionic diffusivity profile around a polymer in the presence of salt, with that of the chain's relevant length-scales (see SI Fig.4). We found  that the effective length scales demarcating adsorbed-layer and diffuse layer around a flexible PE are comparable or even larger than the chain's radius of gyration. This is a result of strong coupling of ions with chain's conformational degree of freedom, unlike the case of rod-like polymer, where the ionic environment is primarily dictated by the Debye length $\lambda_D$.

 \begin{figure}
\includegraphics*[width=\linewidth]{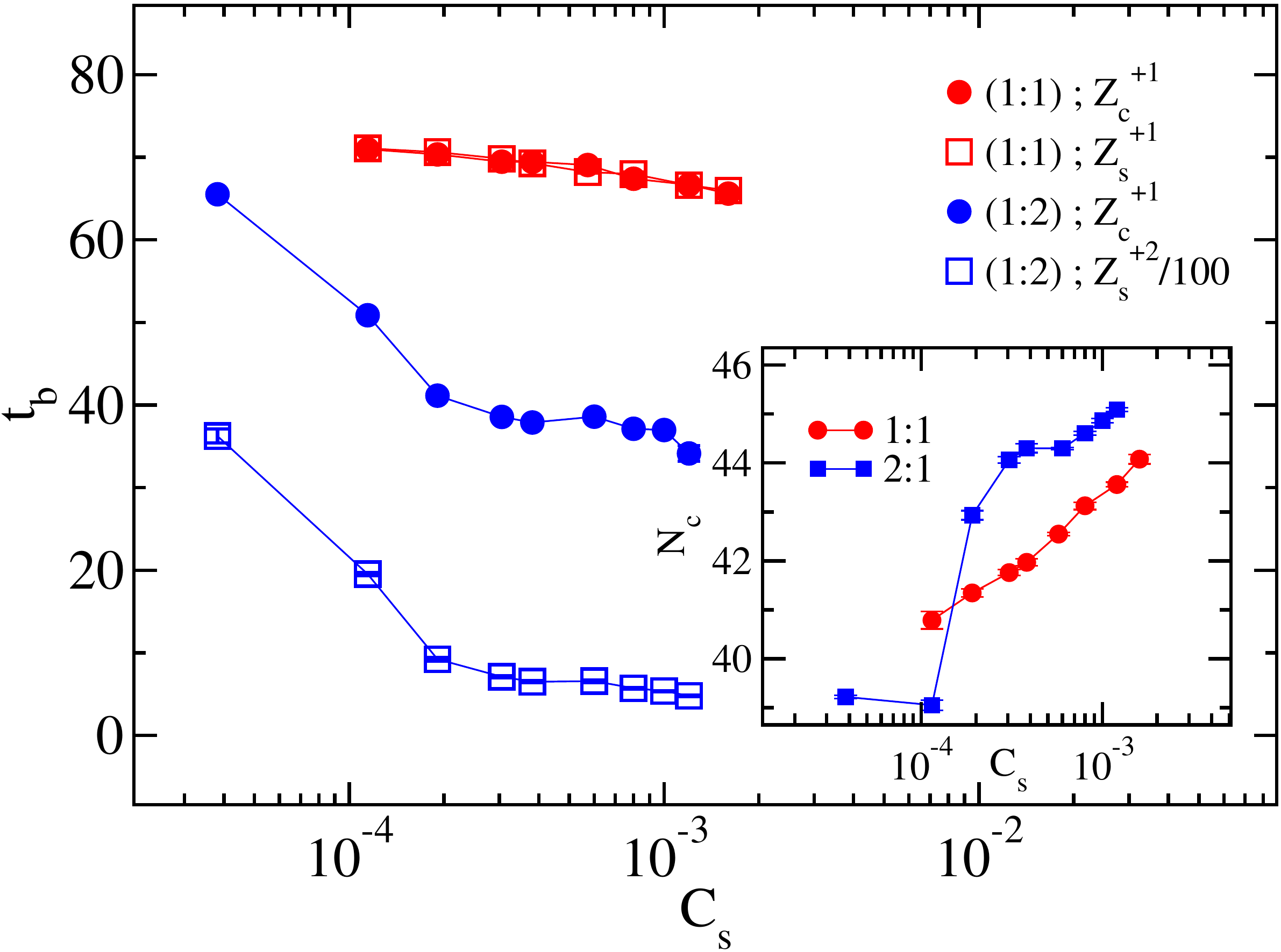}
\caption{ The association time $t_b$ of ions within a chosen cutoff of $\rho_c=1.4$ as a function of salt concentration $C_s$  for chain length $N_m=100$ with $C_m\approx0.0004$. Open symbols pertain to cationic salt ions of valency +z and closed symbols pertains to cationic counterions of valency +1.
The number of condensed ions (including counterions and salt cations) within the Stern layer $\rho_c<1$, on the chain for varying salt concentrations $C_s$ ,where red circles are for 1:1 salt and blue squares represent 2:1 salt case.}
\label{fig:salt_binding_time}
\end{figure}

The ion association time with the chain is presented  in Fig.~\ref{fig:salt_binding_time} for a range of varying salt concentrations $C_s$. For the monovalent salt (1:1), both $Z_c^{+1}$ counterions and $Z_s^{+1}$ salt cations exhibit similar decrease in binding time with $C_s$\cite{jia2012dynamic}. This overlap is again a result of indistinguishability of counterions and cations in monovalent salt.  Similar, decrease in binding time for both salt cation and counterion is observed in the divalent case. Here, the binding time of  $Z_s^{+2}$, is exceptionally larger than its monovalent counterpart i.e., $Z_c^{+1}$. Behaviour of $Z_s^{+2}$ is shown in square open symbols re-scaled by a factor of 1/100 for the sake of representation. This difference can be attributed to the stronger $Z_s^{+2}$-monomer coupling ($\Gamma=6$) than the  $Z_c^{+1}$-monomer coupling ($\Gamma=3$), where the binding time exhibits an exponential dependence on $\Gamma$ (see sec.~ \ref{sec:binding_time}). This  variation in ion binding time with $C_s$, qualitatively resembles the change seen in number of condensed ions $N_c$ on the chain, which is shown in Fig.~\ref{fig:salt_binding_time} (inset). For  1:1 salt, the binding time  $t_b$ exhibits a weak linear dependence on the number of absorbed ions i.e.  $t_b\sim1/N_c$\cite{jia2012dynamic}. This can be attributed to the enhanced exchange rate of ions across the Stern layer, due to the excess ions present at higher concentration, leading to a decrease in average binding time\cite{shi2021counterion}. Further, in  the case of divalent  salt, the binding time of ions (both $Z_s^{+2}$ and $Z_c^{+1}$) exhibits a remarkably strong dependence on the number of adsorbed ions. 
With the change in concentration from $0.00004$ to $0.001$ an increment of roughly 1.1 times in the number of condensed ions leads to the reduction in binding time of the divalent cations to $\approx 1/8^{th}$ of its value. While, $t_b$ of the monovalent counterions reduces to nearly $1/5$ times its value.


\section{Discussion and Conclusion}
We employed  molecular dynamics simulations to carry out a comprehensive study of the dynamics of ions in the proximity of a PE chain under varying physiological conditions. These includes the solvent quality, molecular weight of the chain, valency of added salt and its concentration. Additionally, our  model incorporates solvent mediated  hydrodynamic interactions  via MPC simulation. This work elucidates the dynamic transport coefficient of counterions and  demarcates various regimes of chain-ion coupling. A detail study of  adsorption-desorption  time-scales of ions,  evolution of effective charge fluctuations across monomeric sites, and the effect of added salt on the ion dynamics are also presented. 


An ion in general under the interactive influence of the chain exhibits a sub-diffusive behaviour. Especially, the one cohabiting the chain's backbone, inherits its  diffusive timescales as it gets dragged with the chain\cite{Frank_EPL_2008,karatasos2009dynamics,grass2009polyelectrolytes}. However, our findings suggest that if looked from the chain's frame of reference, these counterions executes an effective diffusion $\langle \delta R^2\rangle \sim t^{\delta};~ \delta\approx 0.9$  throughout the chain-counterion complex. The ion diffusivity is predominantly oriented along the chain, with a comparatively slow diffusive mode $\langle \delta \rho^2\rangle \sim t^{0.4}$ (for $\Gamma=2.8$) seen perpendicularly outwards from the chain. This suggests that an ion even when strongly adhered to the chain, still exercises its translational entropy by hopping along the chain sites, reflected as enhanced residence time of ions on the chain\cite{cui2007counterion,prabhu2004counterion,morfin2004adsorption}.  
These short scale yet pronounced local ion diffusivity is especially conducive in demarcating various regimes of chain-counterion coupling. We have characterised ion’s effective-diffusivity in terms of  both chain’s COM and its segment. The results obtained from the latter suggests that  in the region, where the counterions directly grazes the chain backbone (Stern layer), the ion diffusivity is the lowest followed by a continuous accent in the diffusivity (diffuse layer), before reaching a plateau of free ion diffusivity into the solution. The characterization of diffusivity as a function of its distance from the chain backbone confers a  narrow window of approximately $0<\rho<2$ encapsulating the whole EDL. However,  the estimated diffusivity as a function of its distance from COM unveils another important aspect. Any ion navigating in the neighbourhood of a flexible chain is often frequented with fluctuating chain segments. Hence, in effect  these ions possess  an average diffusivity within chain's conformational boundaries; thereof, the variation from the lowest diffusivity to the highest diffusivity of free ions happens over scales of chain dimension.

 Mentionably,  different time scales of ion diffusion across the PE chain associated with both condensed and free ion, as well as the motion of ions between different PEs is retrieved in dielectric spectroscopy experiments\cite{fischer2013low,kim2015barrier}. These lead to various characteristic spectral contributions. We leverage our access to the distinct length scales of ion association found around the PE chain and varying ion diffusivity across regions to estimate few of these ion relaxation modes (see SI sec. II).   The timescales for the condensed ion $\tau_c=\frac{{R_{c_1}}^2}{D_R(0)}$, and  the free ion  $\tau_{f}=\frac{{(L-R_{c2})}^2}{D_0}$ retrieved here are within an order different and spans the MHz range (HF),  hence we can assert that the relaxation modes of these ions are not sufficient to account for the gap between the high frequency (MHz) and low frequency (KHz) modes found in the experiments\cite{fischer2013low,bordi2004dielectric}.  More-importantly, the associated  time scales of the condensed ions  are smaller than the free ions ($\tau_f>\tau_c$) over specified length scales, and is consistent with the previous work's assertion\cite{fischer2013low}  (see SI sec. II). 


The lowest diffusivity of ions within the chain boundaries ($R<R_g$), exhibits an exponential dependence on the electrostatic interaction strength ($\Gamma$). Remarkably, unlike the case of absolute diffusivity of bound ions, the ion's local diffusivity w.r.t. the chain is independent of molecular weight variation. This is reminiscent of the independent ionic fluctuation at scales much smaller than the polymer's diffusive time scale. Further, these ions remain entraped within the spatial volume spanned by the chain for prolonged times \cite{lifson1962self}, fluctuating back and forth between events of ion adsorption and desorption. Due to this spatial constriction imposed by the fluctuating chain, events pertaining to complete decoupling of counterions and subsequent free ion release are rare. As a result, the average residence time of an ion grazing the chain backbone exhibits an exponential dependence on interactive strength $\Gamma$, directly stemming from the Arrhenius behavior $t_b \sim {\exp ({{-U_c}/{k_BT}})}$, where the electrostatic energy  $U_c 	\propto \Gamma$. Conversely, the definite span an ion stays free before getting reabsorbed remains unaltered with $\Gamma$, due to screening effectuated beyond the Stern layer.

Across the binding layer a dynamic ionic equilibrium is maintained such that the effective charge of the chain fluctuates around a constant value. This fluctuation in the effective charge per site exhibits short range negative correlations in space. The ion flow near the chain is precursor to these correlations as it causes the concomittance of an ion rich and an ion deficit site on the chain, which then leads to a  segmental coalescing and resultant chain shrinkage\cite{kirkwood1952forces}. 


Further, we have shown that higher valency induces stronger ion-monomer coupling ($\Gamma=z_il_B/l_0$), resulting in exceptionally slow diffusivity of salt ions compared to their monovalent counterparts near the chain backbone. Higher valent ions also effectuate mechanism of ion bridging among monomeric sites, which further retards their movement near the chain. Notably, it has been shown that the concentration of salt does not impart any change in the effective diffusivity of ions relative to the chain. However, the local concentration of ions near the chain exhibits a profound influence on the binding timescales of an ion, unlike the diffusivity\cite{shi2021counterion}. Especially for the higher valent salt, a small addition to the local ion population on the chain  substantially decreases the binding time of the ions. 

 Our results especially  the sub-diffusive nature and slow local effective diffusivity of adsorbed ions, will be helpful to provide insight into ionic conductivity near the chain\cite{Frank_EPL_2008,Holm_PRL_2014,Grass_PRL_2008}, that has implications in the understanding of ionic currents in nano-pore experiments where ions are dragged by the PE chain\cite{Dekker_ACSNANO_2012,Muthu_Book_2011}. The estimated binding time of the ions can be directly related to the rate constant of ion exchange for poly-acids, where it is a crucial ingredient in defining the charge-regularization or effective charge of the chain\cite{shi2021counterion,Muthu_Book_2011,muthukumar201750th,Oosawa_1971}. Moreover, these ion association timescales are important in dielectric dispersion experiments\cite{katsumoto2007dielectric}, where they closely affect the chain's relaxation under fields\cite{fixman1980charged,rau1981polarization,washizu2006electric} and other dielectric properties. {\cblue Following this, an important extension of this work would be to employ the ion characterization done here to decipher distinct modes of ion relaxation picked up in dielectric spectroscopy experiments of PE solutions\cite{fischer2013low,kim2015barrier,bordi2004dielectric}}. Further, the influence of weak electric field on aspects like the effective charge correlations, transport coefficients, and binding time of ions is also worth considering in future work and is crucial in the constructive build up of the present subject. {\cblue Additionally, an extension of the present work from the coarse-grained simulations to the atomistic approach would further add to the understanding of the ion dynamics around a PE, where such models intrinsically takes into account the dielectric variation near the chain\cite{smiatek2020theoretical}.  }

\section{Acknowledgement}
This work received financial support from the DST SERB Grant No.  CRG/2020/000661.  High-performance computing facility provided at IISER Bhopal and Paramshivay NSM facility at IIT-BHU  are also acknowledged.

\bibliography{main} 

\begin{thebibliography}{79}%
\makeatletter
\providecommand \@ifxundefined [1]{%
 \@ifx{#1\undefined}
}%
\providecommand \@ifnum [1]{%
 \ifnum #1\expandafter \@firstoftwo
 \else \expandafter \@secondoftwo
 \fi
}%
\providecommand \@ifx [1]{%
 \ifx #1\expandafter \@firstoftwo
 \else \expandafter \@secondoftwo
 \fi
}%
\providecommand \natexlab [1]{#1}%
\providecommand \enquote  [1]{``#1''}%
\providecommand \bibnamefont  [1]{#1}%
\providecommand \bibfnamefont [1]{#1}%
\providecommand \citenamefont [1]{#1}%
\providecommand \href@noop [0]{\@secondoftwo}%
\providecommand \href [0]{\begingroup \@sanitize@url \@href}%
\providecommand \@href[1]{\@@startlink{#1}\@@href}%
\providecommand \@@href[1]{\endgroup#1\@@endlink}%
\providecommand \@sanitize@url [0]{\catcode `\\12\catcode `\$12\catcode
  `\&12\catcode `\#12\catcode `\^12\catcode `\_12\catcode `\%12\relax}%
\providecommand \@@startlink[1]{}%
\providecommand \@@endlink[0]{}%
\providecommand \url  [0]{\begingroup\@sanitize@url \@url }%
\providecommand \@url [1]{\endgroup\@href {#1}{\urlprefix }}%
\providecommand \urlprefix  [0]{URL }%
\providecommand \Eprint [0]{\href }%
\providecommand \doibase [0]{https://doi.org/}%
\providecommand \selectlanguage [0]{\@gobble}%
\providecommand \bibinfo  [0]{\@secondoftwo}%
\providecommand \bibfield  [0]{\@secondoftwo}%
\providecommand \translation [1]{[#1]}%
\providecommand \BibitemOpen [0]{}%
\providecommand \bibitemStop [0]{}%
\providecommand \bibitemNoStop [0]{.\EOS\space}%
\providecommand \EOS [0]{\spacefactor3000\relax}%
\providecommand \BibitemShut  [1]{\csname bibitem#1\endcsname}%
\let\auto@bib@innerbib\@empty
\bibitem [{\citenamefont {Muthukumar}(2017)}]{muthukumar201750th}%
  \BibitemOpen
  \bibfield  {author} {\bibinfo {author} {\bibfnamefont {M.}~\bibnamefont
  {Muthukumar}},\ }\bibfield  {title} {\bibinfo {title} {50th anniversary
  perspective: A perspective on polyelectrolyte solutions},\ }\href@noop {}
  {\bibfield  {journal} {\bibinfo  {journal} {Macromolecules}\ }\textbf
  {\bibinfo {volume} {50}},\ \bibinfo {pages} {9528} (\bibinfo {year}
  {2017})}\BibitemShut {NoStop}%
\bibitem [{\citenamefont {Muthukumar}(2011)}]{Muthu_Book_2011}%
  \BibitemOpen
  \bibfield  {author} {\bibinfo {author} {\bibfnamefont {M.}~\bibnamefont
  {Muthukumar}},\ }\href@noop {} {\emph {\bibinfo {title} {Polymer
  translocation}}}\ (\bibinfo  {publisher} {Taylor \& Francis US},\ \bibinfo
  {year} {2011})\BibitemShut {NoStop}%
\bibitem [{\citenamefont {Khokhlov}\ and\ \citenamefont
  {Khachaturian}(1982)}]{khokhlov1982theory}%
  \BibitemOpen
  \bibfield  {author} {\bibinfo {author} {\bibfnamefont {A.}~\bibnamefont
  {Khokhlov}}\ and\ \bibinfo {author} {\bibfnamefont {K.}~\bibnamefont
  {Khachaturian}},\ }\bibfield  {title} {\bibinfo {title} {On the theory of
  weakly charged polyelectrolytes},\ }\href@noop {} {\bibfield  {journal}
  {\bibinfo  {journal} {Polymer}\ }\textbf {\bibinfo {volume} {23}},\ \bibinfo
  {pages} {1742} (\bibinfo {year} {1982})}\BibitemShut {NoStop}%
\bibitem [{\citenamefont {Dobrynin}\ \emph {et~al.}(1995)\citenamefont
  {Dobrynin}, \citenamefont {Colby},\ and\ \citenamefont
  {Rubinstein}}]{dobrynin1995scaling}%
  \BibitemOpen
  \bibfield  {author} {\bibinfo {author} {\bibfnamefont {A.~V.}\ \bibnamefont
  {Dobrynin}}, \bibinfo {author} {\bibfnamefont {R.~H.}\ \bibnamefont
  {Colby}},\ and\ \bibinfo {author} {\bibfnamefont {M.}~\bibnamefont
  {Rubinstein}},\ }\bibfield  {title} {\bibinfo {title} {Scaling theory of
  polyelectrolyte solutions},\ }\href@noop {} {\bibfield  {journal} {\bibinfo
  {journal} {Macromolecules}\ }\textbf {\bibinfo {volume} {28}},\ \bibinfo
  {pages} {1859} (\bibinfo {year} {1995})}\BibitemShut {NoStop}%
\bibitem [{\citenamefont {Holm}\ \emph {et~al.}(2004)\citenamefont {Holm},
  \citenamefont {Joanny}, \citenamefont {Kremer}, \citenamefont {Netz},
  \citenamefont {Reineker}, \citenamefont {Seidel}, \citenamefont {Vilgis},\
  and\ \citenamefont {Winkler}}]{holm2004polyelectrolyte}%
  \BibitemOpen
  \bibfield  {author} {\bibinfo {author} {\bibfnamefont {C.}~\bibnamefont
  {Holm}}, \bibinfo {author} {\bibfnamefont {J.}~\bibnamefont {Joanny}},
  \bibinfo {author} {\bibfnamefont {K.}~\bibnamefont {Kremer}}, \bibinfo
  {author} {\bibfnamefont {R.}~\bibnamefont {Netz}}, \bibinfo {author}
  {\bibfnamefont {P.}~\bibnamefont {Reineker}}, \bibinfo {author}
  {\bibfnamefont {C.}~\bibnamefont {Seidel}}, \bibinfo {author} {\bibfnamefont
  {T.~A.}\ \bibnamefont {Vilgis}},\ and\ \bibinfo {author} {\bibfnamefont
  {R.}~\bibnamefont {Winkler}},\ }\bibfield  {title} {\bibinfo {title}
  {Polyelectrolyte theory},\ }\href@noop {} {\bibfield  {journal} {\bibinfo
  {journal} {Polyelectrolytes with defined molecular architecture II}\ ,\
  \bibinfo {pages} {67}} (\bibinfo {year} {2004})}\BibitemShut {NoStop}%
\bibitem [{\citenamefont {Netz}\ and\ \citenamefont
  {Andelman}(2003)}]{netz2003neutral}%
  \BibitemOpen
  \bibfield  {author} {\bibinfo {author} {\bibfnamefont {R.~R.}\ \bibnamefont
  {Netz}}\ and\ \bibinfo {author} {\bibfnamefont {D.}~\bibnamefont
  {Andelman}},\ }\bibfield  {title} {\bibinfo {title} {Neutral and charged
  polymers at interfaces},\ }\href@noop {} {\bibfield  {journal} {\bibinfo
  {journal} {Physics reports}\ }\textbf {\bibinfo {volume} {380}},\ \bibinfo
  {pages} {1} (\bibinfo {year} {2003})}\BibitemShut {NoStop}%
\bibitem [{\citenamefont {Barrat}\ and\ \citenamefont
  {Joanny}(1997)}]{barrat1997theory}%
  \BibitemOpen
  \bibfield  {author} {\bibinfo {author} {\bibfnamefont {J.-L.}\ \bibnamefont
  {Barrat}}\ and\ \bibinfo {author} {\bibfnamefont {J.-F.}\ \bibnamefont
  {Joanny}},\ }\bibfield  {title} {\bibinfo {title} {Theory of polyelectrolyte
  solutions},\ }\href@noop {} {\bibfield  {journal} {\bibinfo  {journal}
  {Advances in chemical physics}\ }\textbf {\bibinfo {volume} {94}},\ \bibinfo
  {pages} {1} (\bibinfo {year} {1997})}\BibitemShut {NoStop}%
\bibitem [{\citenamefont {Oosawa}(1971)}]{Oosawa_1971}%
  \BibitemOpen
  \bibfield  {author} {\bibinfo {author} {\bibfnamefont {F.}~\bibnamefont
  {Oosawa}},\ }\href@noop {} {\emph {\bibinfo {title} {Polyelectrolytes}}}\
  (\bibinfo  {publisher} {Marcel Dekker},\ \bibinfo {address} {New York},\
  \bibinfo {year} {1971})\BibitemShut {NoStop}%
\bibitem [{\citenamefont {Winkler}\ \emph {et~al.}(1998)\citenamefont
  {Winkler}, \citenamefont {Gold},\ and\ \citenamefont
  {Reineker}}]{Winkler_PRL_1998}%
  \BibitemOpen
  \bibfield  {author} {\bibinfo {author} {\bibfnamefont {R.~G.}\ \bibnamefont
  {Winkler}}, \bibinfo {author} {\bibfnamefont {M.}~\bibnamefont {Gold}},\ and\
  \bibinfo {author} {\bibfnamefont {P.}~\bibnamefont {Reineker}},\ }\bibfield
  {title} {\bibinfo {title} {Collapse of polyelectrolyte macromolecules by
  counterion condensation and ion pair formation: a molecular dynamics
  simulation study},\ }\href@noop {} {\bibfield  {journal} {\bibinfo  {journal}
  {Physical review letters}\ }\textbf {\bibinfo {volume} {80}},\ \bibinfo
  {pages} {3731} (\bibinfo {year} {1998})}\BibitemShut {NoStop}%
\bibitem [{\citenamefont {Muthukumar}\ and\ \citenamefont
  {Kong}(2006)}]{muthukumar2006simulation}%
  \BibitemOpen
  \bibfield  {author} {\bibinfo {author} {\bibfnamefont {M.}~\bibnamefont
  {Muthukumar}}\ and\ \bibinfo {author} {\bibfnamefont {C.}~\bibnamefont
  {Kong}},\ }\bibfield  {title} {\bibinfo {title} {Simulation of polymer
  translocation through protein channels},\ }\href@noop {} {\bibfield
  {journal} {\bibinfo  {journal} {Proceedings of the National Academy of
  Sciences}\ }\textbf {\bibinfo {volume} {103}},\ \bibinfo {pages} {5273}
  (\bibinfo {year} {2006})}\BibitemShut {NoStop}%
\bibitem [{\citenamefont {Grass}\ \emph {et~al.}(2008)\citenamefont {Grass},
  \citenamefont {B\"ohme}, \citenamefont {Scheler}, \citenamefont {Cottet},\
  and\ \citenamefont {Holm}}]{Grass_PRL_2008}%
  \BibitemOpen
  \bibfield  {author} {\bibinfo {author} {\bibfnamefont {K.}~\bibnamefont
  {Grass}}, \bibinfo {author} {\bibfnamefont {U.}~\bibnamefont {B\"ohme}},
  \bibinfo {author} {\bibfnamefont {U.}~\bibnamefont {Scheler}}, \bibinfo
  {author} {\bibfnamefont {H.}~\bibnamefont {Cottet}},\ and\ \bibinfo {author}
  {\bibfnamefont {C.}~\bibnamefont {Holm}},\ }\bibfield  {title} {\bibinfo
  {title} {Importance of hydrodynamic shielding for the dynamic behavior of
  short polyelectrolyte chains},\ }\href@noop {} {\bibfield  {journal}
  {\bibinfo  {journal} {Phys. Rev. Lett.}\ }\textbf {\bibinfo {volume} {100}},\
  \bibinfo {pages} {096104} (\bibinfo {year} {2008})}\BibitemShut {NoStop}%
\bibitem [{\citenamefont {Hsiao}\ and\ \citenamefont
  {Luijten}(2006)}]{hsiao2006salt}%
  \BibitemOpen
  \bibfield  {author} {\bibinfo {author} {\bibfnamefont {P.-Y.}\ \bibnamefont
  {Hsiao}}\ and\ \bibinfo {author} {\bibfnamefont {E.}~\bibnamefont
  {Luijten}},\ }\bibfield  {title} {\bibinfo {title} {Salt-induced collapse and
  reexpansion of highly charged flexible polyelectrolytes},\ }\href@noop {}
  {\bibfield  {journal} {\bibinfo  {journal} {Physical review letters}\
  }\textbf {\bibinfo {volume} {97}},\ \bibinfo {pages} {148301} (\bibinfo
  {year} {2006})}\BibitemShut {NoStop}%
\bibitem [{\citenamefont {Kundagrami}\ and\ \citenamefont
  {Muthukumar}(2008)}]{kundagrami2008theory}%
  \BibitemOpen
  \bibfield  {author} {\bibinfo {author} {\bibfnamefont {A.}~\bibnamefont
  {Kundagrami}}\ and\ \bibinfo {author} {\bibfnamefont {M.}~\bibnamefont
  {Muthukumar}},\ }\bibfield  {title} {\bibinfo {title} {Theory of competitive
  counterion adsorption on flexible polyelectrolytes: Divalent salts},\
  }\href@noop {} {\bibfield  {journal} {\bibinfo  {journal} {The Journal of
  chemical physics}\ }\textbf {\bibinfo {volume} {128}},\ \bibinfo {pages}
  {244901} (\bibinfo {year} {2008})}\BibitemShut {NoStop}%
\bibitem [{\citenamefont {Lo}\ \emph {et~al.}(2008)\citenamefont {Lo},
  \citenamefont {Khusid},\ and\ \citenamefont {Koplik}}]{lo2008dynamical}%
  \BibitemOpen
  \bibfield  {author} {\bibinfo {author} {\bibfnamefont {T.~S.}\ \bibnamefont
  {Lo}}, \bibinfo {author} {\bibfnamefont {B.}~\bibnamefont {Khusid}},\ and\
  \bibinfo {author} {\bibfnamefont {J.}~\bibnamefont {Koplik}},\ }\bibfield
  {title} {\bibinfo {title} {Dynamical clustering of counterions on flexible
  polyelectrolytes},\ }\href@noop {} {\bibfield  {journal} {\bibinfo  {journal}
  {Physical review letters}\ }\textbf {\bibinfo {volume} {100}},\ \bibinfo
  {pages} {128301} (\bibinfo {year} {2008})}\BibitemShut {NoStop}%
\bibitem [{\citenamefont {Radhakrishnan}\ and\ \citenamefont
  {Singh}(2021)}]{radhakrishnan2021collapse}%
  \BibitemOpen
  \bibfield  {author} {\bibinfo {author} {\bibfnamefont {K.}~\bibnamefont
  {Radhakrishnan}}\ and\ \bibinfo {author} {\bibfnamefont {S.~P.}\ \bibnamefont
  {Singh}},\ }\bibfield  {title} {\bibinfo {title} {Collapse of a confined
  polyelectrolyte chain under an ac electric field},\ }\href@noop {} {\bibfield
   {journal} {\bibinfo  {journal} {Macromolecules}\ }\textbf {\bibinfo {volume}
  {54}},\ \bibinfo {pages} {7998} (\bibinfo {year} {2021})}\BibitemShut
  {NoStop}%
\bibitem [{\citenamefont {Bordi}\ \emph {et~al.}(2004)\citenamefont {Bordi},
  \citenamefont {Cametti},\ and\ \citenamefont {Colby}}]{bordi2004dielectric}%
  \BibitemOpen
  \bibfield  {author} {\bibinfo {author} {\bibfnamefont {F.}~\bibnamefont
  {Bordi}}, \bibinfo {author} {\bibfnamefont {C.}~\bibnamefont {Cametti}},\
  and\ \bibinfo {author} {\bibfnamefont {R.}~\bibnamefont {Colby}},\ }\bibfield
   {title} {\bibinfo {title} {Dielectric spectroscopy and conductivity of
  polyelectrolyte solutions},\ }\href@noop {} {\bibfield  {journal} {\bibinfo
  {journal} {Journal of Physics: Condensed Matter}\ }\textbf {\bibinfo {volume}
  {16}},\ \bibinfo {pages} {R1423} (\bibinfo {year} {2004})}\BibitemShut
  {NoStop}%
\bibitem [{\citenamefont {Liu}\ and\ \citenamefont
  {Muthukumar}(2002{\natexlab{a}})}]{Liu_JCP_2002}%
  \BibitemOpen
  \bibfield  {author} {\bibinfo {author} {\bibfnamefont {S.}~\bibnamefont
  {Liu}}\ and\ \bibinfo {author} {\bibfnamefont {M.}~\bibnamefont
  {Muthukumar}},\ }\bibfield  {title} {\bibinfo {title} {Langevin dynamics
  simulation of counterion distribution around isolated flexible
  polyelectrolyte chains},\ }\href@noop {} {\bibfield  {journal} {\bibinfo
  {journal} {The Journal of chemical physics}\ }\textbf {\bibinfo {volume}
  {116}},\ \bibinfo {pages} {9975} (\bibinfo {year}
  {2002}{\natexlab{a}})}\BibitemShut {NoStop}%
\bibitem [{\citenamefont {Golestanian}\ \emph {et~al.}(1999)\citenamefont
  {Golestanian}, \citenamefont {Kardar},\ and\ \citenamefont
  {Liverpool}}]{golestanian1999collapse}%
  \BibitemOpen
  \bibfield  {author} {\bibinfo {author} {\bibfnamefont {R.}~\bibnamefont
  {Golestanian}}, \bibinfo {author} {\bibfnamefont {M.}~\bibnamefont
  {Kardar}},\ and\ \bibinfo {author} {\bibfnamefont {T.~B.}\ \bibnamefont
  {Liverpool}},\ }\bibfield  {title} {\bibinfo {title} {Collapse of stiff
  polyelectrolytes due to counterion fluctuations},\ }\href@noop {} {\bibfield
  {journal} {\bibinfo  {journal} {Physical review letters}\ }\textbf {\bibinfo
  {volume} {82}},\ \bibinfo {pages} {4456} (\bibinfo {year}
  {1999})}\BibitemShut {NoStop}%
\bibitem [{\citenamefont {Radhakrishnan}\ and\ \citenamefont
  {Singh}(2019{\natexlab{a}})}]{Keerthi_Singh}%
  \BibitemOpen
  \bibfield  {author} {\bibinfo {author} {\bibfnamefont {K.}~\bibnamefont
  {Radhakrishnan}}\ and\ \bibinfo {author} {\bibfnamefont {S.~P.}\ \bibnamefont
  {Singh}},\ }\bibfield  {title} {\bibinfo {title} {Force driven transition of
  a globular polyelectrolyte},\ }\href@noop {} {\bibfield  {journal} {\bibinfo
  {journal} {The Journal of chemical physics}\ }\textbf {\bibinfo {volume}
  {151}},\ \bibinfo {pages} {174902} (\bibinfo {year}
  {2019}{\natexlab{a}})}\BibitemShut {NoStop}%
\bibitem [{\citenamefont {Muthukumar}(1997)}]{Muthu_JCP_1997}%
  \BibitemOpen
  \bibfield  {author} {\bibinfo {author} {\bibfnamefont {M.}~\bibnamefont
  {Muthukumar}},\ }\href@noop {} {\bibfield  {journal} {\bibinfo  {journal} {J.
  Chem. Phys.}\ }\textbf {\bibinfo {volume} {107}},\ \bibinfo {pages} {2619}
  (\bibinfo {year} {1997})}\BibitemShut {NoStop}%
\bibitem [{\citenamefont {Liu}\ \emph {et~al.}(2003)\citenamefont {Liu},
  \citenamefont {Ghosh},\ and\ \citenamefont
  {Muthukumar}}]{liu2003polyelectrolyte}%
  \BibitemOpen
  \bibfield  {author} {\bibinfo {author} {\bibfnamefont {S.}~\bibnamefont
  {Liu}}, \bibinfo {author} {\bibfnamefont {K.}~\bibnamefont {Ghosh}},\ and\
  \bibinfo {author} {\bibfnamefont {M.}~\bibnamefont {Muthukumar}},\ }\bibfield
   {title} {\bibinfo {title} {Polyelectrolyte solutions with added salt: A
  simulation study},\ }\href@noop {} {\bibfield  {journal} {\bibinfo  {journal}
  {The Journal of chemical physics}\ }\textbf {\bibinfo {volume} {119}},\
  \bibinfo {pages} {1813} (\bibinfo {year} {2003})}\BibitemShut {NoStop}%
\bibitem [{\citenamefont {Gonz{\'a}lez-Mozuelos}\ and\ \citenamefont {De~la
  Cruz}(1995)}]{gonzalez1995ion}%
  \BibitemOpen
  \bibfield  {author} {\bibinfo {author} {\bibfnamefont {P.}~\bibnamefont
  {Gonz{\'a}lez-Mozuelos}}\ and\ \bibinfo {author} {\bibfnamefont {M.~O.}\
  \bibnamefont {De~la Cruz}},\ }\bibfield  {title} {\bibinfo {title} {Ion
  condensation in salt-free dilute polyelectrolyte solutions},\ }\href@noop {}
  {\bibfield  {journal} {\bibinfo  {journal} {The Journal of chemical physics}\
  }\textbf {\bibinfo {volume} {103}},\ \bibinfo {pages} {3145} (\bibinfo {year}
  {1995})}\BibitemShut {NoStop}%
\bibitem [{\citenamefont {Manning}(1979)}]{manning1979counterion}%
  \BibitemOpen
  \bibfield  {author} {\bibinfo {author} {\bibfnamefont {G.~S.}\ \bibnamefont
  {Manning}},\ }\bibfield  {title} {\bibinfo {title} {Counterion binding in
  polyelectrolyte theory},\ }\href@noop {} {\bibfield  {journal} {\bibinfo
  {journal} {Accounts of Chemical Research}\ }\textbf {\bibinfo {volume}
  {12}},\ \bibinfo {pages} {443} (\bibinfo {year} {1979})}\BibitemShut
  {NoStop}%
\bibitem [{\citenamefont {Fixman}(1980)}]{fixman1980charged}%
  \BibitemOpen
  \bibfield  {author} {\bibinfo {author} {\bibfnamefont {M.}~\bibnamefont
  {Fixman}},\ }\bibfield  {title} {\bibinfo {title} {Charged macromolecules in
  external fields. i. the sphere},\ }\href@noop {} {\bibfield  {journal}
  {\bibinfo  {journal} {The Journal of Chemical Physics}\ }\textbf {\bibinfo
  {volume} {72}},\ \bibinfo {pages} {5177} (\bibinfo {year}
  {1980})}\BibitemShut {NoStop}%
\bibitem [{\citenamefont {Rau}\ and\ \citenamefont
  {Charney}(1981)}]{rau1981polarization}%
  \BibitemOpen
  \bibfield  {author} {\bibinfo {author} {\bibfnamefont {D.~C.}\ \bibnamefont
  {Rau}}\ and\ \bibinfo {author} {\bibfnamefont {E.}~\bibnamefont {Charney}},\
  }\bibfield  {title} {\bibinfo {title} {Polarization of the ion atmosphere of
  a charged cylinder},\ }\href@noop {} {\bibfield  {journal} {\bibinfo
  {journal} {Biophysical chemistry}\ }\textbf {\bibinfo {volume} {14}},\
  \bibinfo {pages} {1} (\bibinfo {year} {1981})}\BibitemShut {NoStop}%
\bibitem [{\citenamefont {Washizu}\ and\ \citenamefont
  {Kikuchi}(2006)}]{washizu2006electric}%
  \BibitemOpen
  \bibfield  {author} {\bibinfo {author} {\bibfnamefont {H.}~\bibnamefont
  {Washizu}}\ and\ \bibinfo {author} {\bibfnamefont {K.}~\bibnamefont
  {Kikuchi}},\ }\bibfield  {title} {\bibinfo {title} {Electric polarizability
  of dna in aqueous salt solution},\ }\href@noop {} {\bibfield  {journal}
  {\bibinfo  {journal} {The Journal of Physical Chemistry B}\ }\textbf
  {\bibinfo {volume} {110}},\ \bibinfo {pages} {2855} (\bibinfo {year}
  {2006})}\BibitemShut {NoStop}%
\bibitem [{\citenamefont {Katsumoto}\ \emph {et~al.}(2007)\citenamefont
  {Katsumoto}, \citenamefont {Omori}, \citenamefont {Yamamoto}, \citenamefont
  {Yasuda},\ and\ \citenamefont {Asami}}]{katsumoto2007dielectric}%
  \BibitemOpen
  \bibfield  {author} {\bibinfo {author} {\bibfnamefont {Y.}~\bibnamefont
  {Katsumoto}}, \bibinfo {author} {\bibfnamefont {S.}~\bibnamefont {Omori}},
  \bibinfo {author} {\bibfnamefont {D.}~\bibnamefont {Yamamoto}}, \bibinfo
  {author} {\bibfnamefont {A.}~\bibnamefont {Yasuda}},\ and\ \bibinfo {author}
  {\bibfnamefont {K.}~\bibnamefont {Asami}},\ }\bibfield  {title} {\bibinfo
  {title} {Dielectric dispersion of short single-stranded dna in aqueous
  solutions with and without added salt},\ }\href@noop {} {\bibfield  {journal}
  {\bibinfo  {journal} {Physical Review E}\ }\textbf {\bibinfo {volume} {75}},\
  \bibinfo {pages} {011911} (\bibinfo {year} {2007})}\BibitemShut {NoStop}%
\bibitem [{\citenamefont {Fischer}\ \emph {et~al.}(2008)\citenamefont
  {Fischer}, \citenamefont {Naji},\ and\ \citenamefont
  {Netz}}]{fischer2008salt}%
  \BibitemOpen
  \bibfield  {author} {\bibinfo {author} {\bibfnamefont {S.}~\bibnamefont
  {Fischer}}, \bibinfo {author} {\bibfnamefont {A.}~\bibnamefont {Naji}},\ and\
  \bibinfo {author} {\bibfnamefont {R.~R.}\ \bibnamefont {Netz}},\ }\bibfield
  {title} {\bibinfo {title} {Salt-induced counterion-mobility anomaly in
  polyelectrolyte electrophoresis},\ }\href@noop {} {\bibfield  {journal}
  {\bibinfo  {journal} {Physical review letters}\ }\textbf {\bibinfo {volume}
  {101}},\ \bibinfo {pages} {176103} (\bibinfo {year} {2008})}\BibitemShut
  {NoStop}%
\bibitem [{\citenamefont {Winkler}\ and\ \citenamefont
  {Huang}(2009)}]{Winkler_JCP_2009}%
  \BibitemOpen
  \bibfield  {author} {\bibinfo {author} {\bibfnamefont {R.~G.}\ \bibnamefont
  {Winkler}}\ and\ \bibinfo {author} {\bibfnamefont {C.-C.}\ \bibnamefont
  {Huang}},\ }\bibfield  {title} {\bibinfo {title} {Stress tensors of
  multiparticle collision dynamics fluids},\ }\href@noop {} {\bibfield
  {journal} {\bibinfo  {journal} {J. Chem. Phys.}\ }\textbf {\bibinfo {volume}
  {130}},\ \bibinfo {pages} {074907} (\bibinfo {year} {2009})}\BibitemShut
  {NoStop}%
\bibitem [{\citenamefont {Muthukumar}(2004)}]{muthukumar2004theory}%
  \BibitemOpen
  \bibfield  {author} {\bibinfo {author} {\bibfnamefont {M.}~\bibnamefont
  {Muthukumar}},\ }\bibfield  {title} {\bibinfo {title} {Theory of counter-ion
  condensation on flexible polyelectrolytes: Adsorption mechanism},\
  }\href@noop {} {\bibfield  {journal} {\bibinfo  {journal} {The Journal of
  chemical physics}\ }\textbf {\bibinfo {volume} {120}},\ \bibinfo {pages}
  {9343} (\bibinfo {year} {2004})}\BibitemShut {NoStop}%
\bibitem [{\citenamefont {Liu}\ and\ \citenamefont
  {Muthukumar}(2002{\natexlab{b}})}]{liu2002langevin}%
  \BibitemOpen
  \bibfield  {author} {\bibinfo {author} {\bibfnamefont {S.}~\bibnamefont
  {Liu}}\ and\ \bibinfo {author} {\bibfnamefont {M.}~\bibnamefont
  {Muthukumar}},\ }\bibfield  {title} {\bibinfo {title} {Langevin dynamics
  simulation of counterion distribution around isolated flexible
  polyelectrolyte chains},\ }\href@noop {} {\bibfield  {journal} {\bibinfo
  {journal} {The Journal of chemical physics}\ }\textbf {\bibinfo {volume}
  {116}},\ \bibinfo {pages} {9975} (\bibinfo {year}
  {2002}{\natexlab{b}})}\BibitemShut {NoStop}%
\bibitem [{\citenamefont {Smiatek}(2020)}]{smiatek2020theoretical}%
  \BibitemOpen
  \bibfield  {author} {\bibinfo {author} {\bibfnamefont {J.}~\bibnamefont
  {Smiatek}},\ }\bibfield  {title} {\bibinfo {title} {Theoretical and
  computational insight into solvent and specific ion effects for
  polyelectrolytes: the importance of local molecular interactions},\
  }\href@noop {} {\bibfield  {journal} {\bibinfo  {journal} {Molecules}\
  }\textbf {\bibinfo {volume} {25}},\ \bibinfo {pages} {1661} (\bibinfo {year}
  {2020})}\BibitemShut {NoStop}%
\bibitem [{\citenamefont {Karatasos}\ and\ \citenamefont
  {Krystallis}(2009)}]{karatasos2009dynamics}%
  \BibitemOpen
  \bibfield  {author} {\bibinfo {author} {\bibfnamefont {K.}~\bibnamefont
  {Karatasos}}\ and\ \bibinfo {author} {\bibfnamefont {M.}~\bibnamefont
  {Krystallis}},\ }\bibfield  {title} {\bibinfo {title} {Dynamics of
  counterions in dendrimer polyelectrolyte solutions},\ }\href@noop {}
  {\bibfield  {journal} {\bibinfo  {journal} {The Journal of chemical physics}\
  }\textbf {\bibinfo {volume} {130}},\ \bibinfo {pages} {114903} (\bibinfo
  {year} {2009})}\BibitemShut {NoStop}%
\bibitem [{\citenamefont {Frank}\ and\ \citenamefont
  {Winkler}(2008)}]{Frank_EPL_2008}%
  \BibitemOpen
  \bibfield  {author} {\bibinfo {author} {\bibfnamefont {S.}~\bibnamefont
  {Frank}}\ and\ \bibinfo {author} {\bibfnamefont {R.}~\bibnamefont
  {Winkler}},\ }\bibfield  {title} {\bibinfo {title} {Polyelectrolyte
  electrophoresis: Field effects and hydrodynamic interactions},\ }\href@noop
  {} {\bibfield  {journal} {\bibinfo  {journal} {EPL (Europhysics Letters)}\
  }\textbf {\bibinfo {volume} {83}},\ \bibinfo {pages} {38004} (\bibinfo {year}
  {2008})}\BibitemShut {NoStop}%
\bibitem [{\citenamefont {Singh}\ and\ \citenamefont
  {Muthukumar}(2014)}]{Singh_2014_JCP}%
  \BibitemOpen
  \bibfield  {author} {\bibinfo {author} {\bibfnamefont {S.~P.}\ \bibnamefont
  {Singh}}\ and\ \bibinfo {author} {\bibfnamefont {M.}~\bibnamefont
  {Muthukumar}},\ }\bibfield  {title} {\bibinfo {title} {Electrophoretic
  mobilities of counterions and a polymer in cylindrical pores},\ }\href@noop
  {} {\bibfield  {journal} {\bibinfo  {journal} {The Journal of chemical
  physics}\ }\textbf {\bibinfo {volume} {141}},\ \bibinfo {pages} {09B610\_1}
  (\bibinfo {year} {2014})}\BibitemShut {NoStop}%
\bibitem [{\citenamefont {Chang}\ and\ \citenamefont
  {Yethiraj}(2002)}]{chang2002brownian}%
  \BibitemOpen
  \bibfield  {author} {\bibinfo {author} {\bibfnamefont {R.}~\bibnamefont
  {Chang}}\ and\ \bibinfo {author} {\bibfnamefont {A.}~\bibnamefont
  {Yethiraj}},\ }\bibfield  {title} {\bibinfo {title} {Brownian dynamics
  simulations of salt-free polyelectrolyte solutions},\ }\href@noop {}
  {\bibfield  {journal} {\bibinfo  {journal} {The Journal of chemical physics}\
  }\textbf {\bibinfo {volume} {116}},\ \bibinfo {pages} {5284} (\bibinfo {year}
  {2002})}\BibitemShut {NoStop}%
\bibitem [{\citenamefont {Cui}(2007)}]{cui2007counterion}%
  \BibitemOpen
  \bibfield  {author} {\bibinfo {author} {\bibfnamefont {S.}~\bibnamefont
  {Cui}},\ }\bibfield  {title} {\bibinfo {title} {Counterion-hopping along the
  backbone of single-stranded dna in nanometer pores: A mechanism for current
  conduction},\ }\href@noop {} {\bibfield  {journal} {\bibinfo  {journal}
  {Physical review letters}\ }\textbf {\bibinfo {volume} {98}},\ \bibinfo
  {pages} {138101} (\bibinfo {year} {2007})}\BibitemShut {NoStop}%
\bibitem [{\citenamefont {Prabhu}\ \emph {et~al.}(2004)\citenamefont {Prabhu},
  \citenamefont {Amis}, \citenamefont {Bossev},\ and\ \citenamefont
  {Rosov}}]{prabhu2004counterion}%
  \BibitemOpen
  \bibfield  {author} {\bibinfo {author} {\bibfnamefont {V.~M.}\ \bibnamefont
  {Prabhu}}, \bibinfo {author} {\bibfnamefont {E.~J.}\ \bibnamefont {Amis}},
  \bibinfo {author} {\bibfnamefont {D.~P.}\ \bibnamefont {Bossev}},\ and\
  \bibinfo {author} {\bibfnamefont {N.}~\bibnamefont {Rosov}},\ }\bibfield
  {title} {\bibinfo {title} {Counterion associative behavior with flexible
  polyelectrolytes},\ }\href@noop {} {\bibfield  {journal} {\bibinfo  {journal}
  {The Journal of chemical physics}\ }\textbf {\bibinfo {volume} {121}},\
  \bibinfo {pages} {4424} (\bibinfo {year} {2004})}\BibitemShut {NoStop}%
\bibitem [{\citenamefont {Morfin}\ \emph {et~al.}(2004)\citenamefont {Morfin},
  \citenamefont {Horkay}, \citenamefont {Basser}, \citenamefont {Bley},
  \citenamefont {Hecht}, \citenamefont {Rochas},\ and\ \citenamefont
  {Geissler}}]{morfin2004adsorption}%
  \BibitemOpen
  \bibfield  {author} {\bibinfo {author} {\bibfnamefont {I.}~\bibnamefont
  {Morfin}}, \bibinfo {author} {\bibfnamefont {F.}~\bibnamefont {Horkay}},
  \bibinfo {author} {\bibfnamefont {P.~J.}\ \bibnamefont {Basser}}, \bibinfo
  {author} {\bibfnamefont {F.}~\bibnamefont {Bley}}, \bibinfo {author}
  {\bibfnamefont {A.-M.}\ \bibnamefont {Hecht}}, \bibinfo {author}
  {\bibfnamefont {C.}~\bibnamefont {Rochas}},\ and\ \bibinfo {author}
  {\bibfnamefont {E.}~\bibnamefont {Geissler}},\ }\bibfield  {title} {\bibinfo
  {title} {Adsorption of divalent cations on dna},\ }\href@noop {} {\bibfield
  {journal} {\bibinfo  {journal} {Biophysical journal}\ }\textbf {\bibinfo
  {volume} {87}},\ \bibinfo {pages} {2897} (\bibinfo {year}
  {2004})}\BibitemShut {NoStop}%
\bibitem [{\citenamefont {Yu}\ and\ \citenamefont
  {Iwahara}(2021)}]{yu2021experimental}%
  \BibitemOpen
  \bibfield  {author} {\bibinfo {author} {\bibfnamefont {B.}~\bibnamefont
  {Yu}}\ and\ \bibinfo {author} {\bibfnamefont {J.}~\bibnamefont {Iwahara}},\
  }\bibfield  {title} {\bibinfo {title} {Experimental approaches for
  investigating ion atmospheres around nucleic acids and proteins},\
  }\href@noop {} {\bibfield  {journal} {\bibinfo  {journal} {Computational and
  Structural Biotechnology Journal}\ ,\ } (\bibinfo {year} {2021})}\BibitemShut
  {NoStop}%
\bibitem [{\citenamefont {Shi}\ \emph {et~al.}(2021)\citenamefont {Shi},
  \citenamefont {Peng}, \citenamefont {Yang},\ and\ \citenamefont
  {Zhao}}]{shi2021counterion}%
  \BibitemOpen
  \bibfield  {author} {\bibinfo {author} {\bibfnamefont {Y.}~\bibnamefont
  {Shi}}, \bibinfo {author} {\bibfnamefont {H.}~\bibnamefont {Peng}}, \bibinfo
  {author} {\bibfnamefont {J.}~\bibnamefont {Yang}},\ and\ \bibinfo {author}
  {\bibfnamefont {J.}~\bibnamefont {Zhao}},\ }\bibfield  {title} {\bibinfo
  {title} {Counterion binding dynamics of a polyelectrolyte},\ }\href@noop {}
  {\bibfield  {journal} {\bibinfo  {journal} {Macromolecules}\ }\textbf
  {\bibinfo {volume} {54}},\ \bibinfo {pages} {4926} (\bibinfo {year}
  {2021})}\BibitemShut {NoStop}%
\bibitem [{\citenamefont {Schipper}\ \emph {et~al.}(1997)\citenamefont
  {Schipper}, \citenamefont {Hollander},\ and\ \citenamefont
  {Leyte}}]{schipper1997counterion}%
  \BibitemOpen
  \bibfield  {author} {\bibinfo {author} {\bibfnamefont {F.}~\bibnamefont
  {Schipper}}, \bibinfo {author} {\bibfnamefont {J.}~\bibnamefont
  {Hollander}},\ and\ \bibinfo {author} {\bibfnamefont {J.}~\bibnamefont
  {Leyte}},\ }\bibfield  {title} {\bibinfo {title} {Counterion self-diffusion
  in polyelectrolyte solutions},\ }\href@noop {} {\bibfield  {journal}
  {\bibinfo  {journal} {Journal of Physics: Condensed Matter}\ }\textbf
  {\bibinfo {volume} {9}},\ \bibinfo {pages} {11179} (\bibinfo {year}
  {1997})}\BibitemShut {NoStop}%
\bibitem [{\citenamefont {Torres}\ \emph {et~al.}(2017)\citenamefont {Torres},
  \citenamefont {Bojanich}, \citenamefont {Sanchez-Varretti}, \citenamefont
  {Ramirez-Pastor}, \citenamefont {Quiroga}, \citenamefont {Boeris},\ and\
  \citenamefont {Narambuena}}]{torres2017protonation}%
  \BibitemOpen
  \bibfield  {author} {\bibinfo {author} {\bibfnamefont {P.}~\bibnamefont
  {Torres}}, \bibinfo {author} {\bibfnamefont {L.}~\bibnamefont {Bojanich}},
  \bibinfo {author} {\bibfnamefont {F.}~\bibnamefont {Sanchez-Varretti}},
  \bibinfo {author} {\bibfnamefont {A.~J.}\ \bibnamefont {Ramirez-Pastor}},
  \bibinfo {author} {\bibfnamefont {E.}~\bibnamefont {Quiroga}}, \bibinfo
  {author} {\bibfnamefont {V.}~\bibnamefont {Boeris}},\ and\ \bibinfo {author}
  {\bibfnamefont {C.~F.}\ \bibnamefont {Narambuena}},\ }\bibfield  {title}
  {\bibinfo {title} {Protonation of $\beta$-lactoglobulin in the presence of
  strong polyelectrolyte chains: A study using monte carlo simulation},\
  }\href@noop {} {\bibfield  {journal} {\bibinfo  {journal} {Colloids and
  Surfaces B: Biointerfaces}\ }\textbf {\bibinfo {volume} {160}},\ \bibinfo
  {pages} {161} (\bibinfo {year} {2017})}\BibitemShut {NoStop}%
\bibitem [{\citenamefont {Mason}\ and\ \citenamefont
  {Jensen}(2008)}]{mason2008protein}%
  \BibitemOpen
  \bibfield  {author} {\bibinfo {author} {\bibfnamefont {A.~C.}\ \bibnamefont
  {Mason}}\ and\ \bibinfo {author} {\bibfnamefont {J.~H.}\ \bibnamefont
  {Jensen}},\ }\bibfield  {title} {\bibinfo {title} {Protein--protein binding
  is often associated with changes in protonation state},\ }\href@noop {}
  {\bibfield  {journal} {\bibinfo  {journal} {Proteins: Structure, Function,
  and Bioinformatics}\ }\textbf {\bibinfo {volume} {71}},\ \bibinfo {pages}
  {81} (\bibinfo {year} {2008})}\BibitemShut {NoStop}%
\bibitem [{\citenamefont {Lund}\ and\ \citenamefont
  {J{\"o}nsson}(2005)}]{lund2005charge}%
  \BibitemOpen
  \bibfield  {author} {\bibinfo {author} {\bibfnamefont {M.}~\bibnamefont
  {Lund}}\ and\ \bibinfo {author} {\bibfnamefont {B.}~\bibnamefont
  {J{\"o}nsson}},\ }\bibfield  {title} {\bibinfo {title} {On the charge
  regulation of proteins},\ }\href@noop {} {\bibfield  {journal} {\bibinfo
  {journal} {Biochemistry}\ }\textbf {\bibinfo {volume} {44}},\ \bibinfo
  {pages} {5722} (\bibinfo {year} {2005})}\BibitemShut {NoStop}%
\bibitem [{\citenamefont {Lund}\ and\ \citenamefont
  {J{\"o}nsson}(2013)}]{lund2013charge}%
  \BibitemOpen
  \bibfield  {author} {\bibinfo {author} {\bibfnamefont {M.}~\bibnamefont
  {Lund}}\ and\ \bibinfo {author} {\bibfnamefont {B.}~\bibnamefont
  {J{\"o}nsson}},\ }\bibfield  {title} {\bibinfo {title} {Charge regulation in
  biomolecular solution},\ }\href@noop {} {\bibfield  {journal} {\bibinfo
  {journal} {Quarterly reviews of biophysics}\ }\textbf {\bibinfo {volume}
  {46}},\ \bibinfo {pages} {265} (\bibinfo {year} {2013})}\BibitemShut
  {NoStop}%
\bibitem [{\citenamefont {Koplik}\ and\ \citenamefont
  {Banavar}(1995)}]{Koplik_CFP_1995}%
  \BibitemOpen
  \bibfield  {author} {\bibinfo {author} {\bibfnamefont {J.}~\bibnamefont
  {Koplik}}\ and\ \bibinfo {author} {\bibfnamefont {J.~R.}\ \bibnamefont
  {Banavar}},\ }\bibfield  {title} {\bibinfo {title} {Corner flow in the
  sliding plate problem},\ }\href@noop {} {\bibfield  {journal} {\bibinfo
  {journal} {Physics of Fluids}\ }\textbf {\bibinfo {volume} {7}},\ \bibinfo
  {pages} {3118} (\bibinfo {year} {1995})}\BibitemShut {NoStop}%
\bibitem [{\citenamefont {Angelini}\ \emph {et~al.}(2005)\citenamefont
  {Angelini}, \citenamefont {Sanders}, \citenamefont {Liang}, \citenamefont
  {Wriggers}, \citenamefont {Tang},\ and\ \citenamefont
  {Wong}}]{angelini2005structure}%
  \BibitemOpen
  \bibfield  {author} {\bibinfo {author} {\bibfnamefont {T.~E.}\ \bibnamefont
  {Angelini}}, \bibinfo {author} {\bibfnamefont {L.~K.}\ \bibnamefont
  {Sanders}}, \bibinfo {author} {\bibfnamefont {H.}~\bibnamefont {Liang}},
  \bibinfo {author} {\bibfnamefont {W.}~\bibnamefont {Wriggers}}, \bibinfo
  {author} {\bibfnamefont {J.~X.}\ \bibnamefont {Tang}},\ and\ \bibinfo
  {author} {\bibfnamefont {G.~C.}\ \bibnamefont {Wong}},\ }\bibfield  {title}
  {\bibinfo {title} {Structure and dynamics of condensed multivalent ions
  within polyelectrolyte bundles: a combined x-ray diffraction and solid-state
  nmr study},\ }\href@noop {} {\bibfield  {journal} {\bibinfo  {journal}
  {Journal of Physics: Condensed Matter}\ }\textbf {\bibinfo {volume} {17}},\
  \bibinfo {pages} {S1123} (\bibinfo {year} {2005})}\BibitemShut {NoStop}%
\bibitem [{\citenamefont {Kirkwood}\ and\ \citenamefont
  {Shumaker}(1952)}]{kirkwood1952forces}%
  \BibitemOpen
  \bibfield  {author} {\bibinfo {author} {\bibfnamefont {J.~G.}\ \bibnamefont
  {Kirkwood}}\ and\ \bibinfo {author} {\bibfnamefont {J.~B.}\ \bibnamefont
  {Shumaker}},\ }\bibfield  {title} {\bibinfo {title} {Forces between protein
  molecules in solution arising from fluctuations in proton charge and
  configuration},\ }\href@noop {} {\bibfield  {journal} {\bibinfo  {journal}
  {Proceedings of the National Academy of Sciences of the United States of
  America}\ }\textbf {\bibinfo {volume} {38}},\ \bibinfo {pages} {863}
  (\bibinfo {year} {1952})}\BibitemShut {NoStop}%
\bibitem [{\citenamefont {Blanco}\ \emph {et~al.}(2019)\citenamefont {Blanco},
  \citenamefont {Madurga}, \citenamefont {Narambuena}, \citenamefont {Mas},\
  and\ \citenamefont {Garc{\'e}s}}]{blanco2019role}%
  \BibitemOpen
  \bibfield  {author} {\bibinfo {author} {\bibfnamefont {P.~M.}\ \bibnamefont
  {Blanco}}, \bibinfo {author} {\bibfnamefont {S.}~\bibnamefont {Madurga}},
  \bibinfo {author} {\bibfnamefont {C.~F.}\ \bibnamefont {Narambuena}},
  \bibinfo {author} {\bibfnamefont {F.}~\bibnamefont {Mas}},\ and\ \bibinfo
  {author} {\bibfnamefont {J.~L.}\ \bibnamefont {Garc{\'e}s}},\ }\bibfield
  {title} {\bibinfo {title} {Role of charge regulation and fluctuations in the
  conformational and mechanical properties of weak flexible polyelectrolytes},\
  }\href@noop {} {\bibfield  {journal} {\bibinfo  {journal} {Polymers}\
  }\textbf {\bibinfo {volume} {11}},\ \bibinfo {pages} {1962} (\bibinfo {year}
  {2019})}\BibitemShut {NoStop}%
\bibitem [{\citenamefont {da~Silva}\ and\ \citenamefont
  {J{\"o}nsson}(2009)}]{da2009polyelectrolyte}%
  \BibitemOpen
  \bibfield  {author} {\bibinfo {author} {\bibfnamefont {F.~L.~B.}\
  \bibnamefont {da~Silva}}\ and\ \bibinfo {author} {\bibfnamefont
  {B.}~\bibnamefont {J{\"o}nsson}},\ }\bibfield  {title} {\bibinfo {title}
  {Polyelectrolyte--protein complexation driven by charge regulation},\
  }\href@noop {} {\bibfield  {journal} {\bibinfo  {journal} {Soft Matter}\
  }\textbf {\bibinfo {volume} {5}},\ \bibinfo {pages} {2862} (\bibinfo {year}
  {2009})}\BibitemShut {NoStop}%
\bibitem [{\citenamefont {Fahrenberger}\ \emph {et~al.}(2015)\citenamefont
  {Fahrenberger}, \citenamefont {Hickey}, \citenamefont {Smiatek},\ and\
  \citenamefont {Holm}}]{fahrenberger2015influence}%
  \BibitemOpen
  \bibfield  {author} {\bibinfo {author} {\bibfnamefont {F.}~\bibnamefont
  {Fahrenberger}}, \bibinfo {author} {\bibfnamefont {O.~A.}\ \bibnamefont
  {Hickey}}, \bibinfo {author} {\bibfnamefont {J.}~\bibnamefont {Smiatek}},\
  and\ \bibinfo {author} {\bibfnamefont {C.}~\bibnamefont {Holm}},\ }\bibfield
  {title} {\bibinfo {title} {The influence of charged-induced variations in the
  local permittivity on the static and dynamic properties of polyelectrolyte
  solutions},\ }\href@noop {} {\bibfield  {journal} {\bibinfo  {journal} {The
  Journal of chemical physics}\ }\textbf {\bibinfo {volume} {143}},\ \bibinfo
  {pages} {243140} (\bibinfo {year} {2015})}\BibitemShut {NoStop}%
\bibitem [{\citenamefont {Deserno}\ and\ \citenamefont
  {Holm}(1998)}]{deserno1998mesh_1}%
  \BibitemOpen
  \bibfield  {author} {\bibinfo {author} {\bibfnamefont {M.}~\bibnamefont
  {Deserno}}\ and\ \bibinfo {author} {\bibfnamefont {C.}~\bibnamefont {Holm}},\
  }\bibfield  {title} {\bibinfo {title} {How to mesh up ewald sums. i. a
  theoretical and numerical comparison of various particle mesh routines},\
  }\href@noop {} {\bibfield  {journal} {\bibinfo  {journal} {The Journal of
  chemical physics}\ }\textbf {\bibinfo {volume} {109}},\ \bibinfo {pages}
  {7678} (\bibinfo {year} {1998})}\BibitemShut {NoStop}%
\bibitem [{\citenamefont {Fahrenberger}\ and\ \citenamefont
  {Holm}(2014)}]{fahrenberger2014computing}%
  \BibitemOpen
  \bibfield  {author} {\bibinfo {author} {\bibfnamefont {F.}~\bibnamefont
  {Fahrenberger}}\ and\ \bibinfo {author} {\bibfnamefont {C.}~\bibnamefont
  {Holm}},\ }\bibfield  {title} {\bibinfo {title} {Computing the coulomb
  interaction in inhomogeneous dielectric media via a local electrostatics
  lattice algorithm},\ }\href@noop {} {\bibfield  {journal} {\bibinfo
  {journal} {Physical Review E}\ }\textbf {\bibinfo {volume} {90}},\ \bibinfo
  {pages} {063304} (\bibinfo {year} {2014})}\BibitemShut {NoStop}%
\bibitem [{\citenamefont {Tyagi}\ \emph {et~al.}(2010)\citenamefont {Tyagi},
  \citenamefont {S{\"u}zen}, \citenamefont {Sega}, \citenamefont {Barbosa},
  \citenamefont {Kantorovich},\ and\ \citenamefont
  {Holm}}]{tyagi2010iterative}%
  \BibitemOpen
  \bibfield  {author} {\bibinfo {author} {\bibfnamefont {S.}~\bibnamefont
  {Tyagi}}, \bibinfo {author} {\bibfnamefont {M.}~\bibnamefont {S{\"u}zen}},
  \bibinfo {author} {\bibfnamefont {M.}~\bibnamefont {Sega}}, \bibinfo {author}
  {\bibfnamefont {M.}~\bibnamefont {Barbosa}}, \bibinfo {author} {\bibfnamefont
  {S.~S.}\ \bibnamefont {Kantorovich}},\ and\ \bibinfo {author} {\bibfnamefont
  {C.}~\bibnamefont {Holm}},\ }\bibfield  {title} {\bibinfo {title} {An
  iterative, fast, linear-scaling method for computing induced charges on
  arbitrary dielectric boundaries},\ }\href@noop {} {\bibfield  {journal}
  {\bibinfo  {journal} {The Journal of chemical physics}\ }\textbf {\bibinfo
  {volume} {132}},\ \bibinfo {pages} {154112} (\bibinfo {year}
  {2010})}\BibitemShut {NoStop}%
\bibitem [{\citenamefont {Kapral}(2008)}]{Kapral_ACP_2008}%
  \BibitemOpen
  \bibfield  {author} {\bibinfo {author} {\bibfnamefont {R.}~\bibnamefont
  {Kapral}},\ }\bibfield  {title} {\bibinfo {title} {Multiparticle collision
  dynamics: {S}imulation of complex systems on mesoscales},\ }\href@noop {}
  {\bibfield  {journal} {\bibinfo  {journal} {Adv. Chem. Phys.}\ }\textbf
  {\bibinfo {volume} {140}},\ \bibinfo {pages} {89} (\bibinfo {year}
  {2008})}\BibitemShut {NoStop}%
\bibitem [{\citenamefont {Gompper}\ \emph {et~al.}(2009)\citenamefont
  {Gompper}, \citenamefont {Ihle}, \citenamefont {Kroll},\ and\ \citenamefont
  {Winkler}}]{Gompper_APS_2009}%
  \BibitemOpen
  \bibfield  {author} {\bibinfo {author} {\bibfnamefont {G.}~\bibnamefont
  {Gompper}}, \bibinfo {author} {\bibfnamefont {T.}~\bibnamefont {Ihle}},
  \bibinfo {author} {\bibfnamefont {D.~M.}\ \bibnamefont {Kroll}},\ and\
  \bibinfo {author} {\bibfnamefont {R.~G.}\ \bibnamefont {Winkler}},\
  }\bibfield  {title} {\bibinfo {title} {Multi-particle collision dynamics: a
  particle-based mesoscale simulation approach to the hydrodynamics of complex
  fluids},\ }\href@noop {} {\bibfield  {journal} {\bibinfo  {journal} {Adv.
  Polym. Sci.}\ }\textbf {\bibinfo {volume} {221}},\ \bibinfo {pages} {1}
  (\bibinfo {year} {2009})}\BibitemShut {NoStop}%
\bibitem [{\citenamefont {Ripoll}\ \emph {et~al.}(2004)\citenamefont {Ripoll},
  \citenamefont {Mussawisade}, \citenamefont {Winkler},\ and\ \citenamefont
  {Gompper}}]{Ripoll04}%
  \BibitemOpen
  \bibfield  {author} {\bibinfo {author} {\bibfnamefont {M.}~\bibnamefont
  {Ripoll}}, \bibinfo {author} {\bibfnamefont {K.}~\bibnamefont {Mussawisade}},
  \bibinfo {author} {\bibfnamefont {R.~G.}\ \bibnamefont {Winkler}},\ and\
  \bibinfo {author} {\bibfnamefont {G.}~\bibnamefont {Gompper}},\ }\bibfield
  {title} {\bibinfo {title} {Low-reynolds-number hydrodynamics of complex
  fluids by multi-particle-collision dynamics},\ }\href@noop {} {\bibfield
  {journal} {\bibinfo  {journal} {Europhys. Lett.}\ }\textbf {\bibinfo {volume}
  {68}},\ \bibinfo {pages} {106} (\bibinfo {year} {2004})}\BibitemShut
  {NoStop}%
\bibitem [{\citenamefont {Malevanets}\ and\ \citenamefont
  {Kapral}(1999)}]{Malevanets_MSM_1999}%
  \BibitemOpen
  \bibfield  {author} {\bibinfo {author} {\bibfnamefont {A.}~\bibnamefont
  {Malevanets}}\ and\ \bibinfo {author} {\bibfnamefont {R.}~\bibnamefont
  {Kapral}},\ }\bibfield  {title} {\bibinfo {title} {Mesoscopic model for
  solvent dynamics},\ }\href@noop {} {\bibfield  {journal} {\bibinfo  {journal}
  {J. Chem. Phys.}\ }\textbf {\bibinfo {volume} {110}},\ \bibinfo {pages}
  {8605} (\bibinfo {year} {1999})}\BibitemShut {NoStop}%
\bibitem [{\citenamefont {Lamura}\ \emph {et~al.}(2001)\citenamefont {Lamura},
  \citenamefont {Gompper}, \citenamefont {Ihle},\ and\ \citenamefont
  {Kroll}}]{Lamura_EPL_2001}%
  \BibitemOpen
  \bibfield  {author} {\bibinfo {author} {\bibfnamefont {A.}~\bibnamefont
  {Lamura}}, \bibinfo {author} {\bibfnamefont {G.}~\bibnamefont {Gompper}},
  \bibinfo {author} {\bibfnamefont {T.}~\bibnamefont {Ihle}},\ and\ \bibinfo
  {author} {\bibfnamefont {D.}~\bibnamefont {Kroll}},\ }\bibfield  {title}
  {\bibinfo {title} {Multi-particle collision dynamics: Flow around a circular
  and a square cylinder},\ }\href@noop {} {\bibfield  {journal} {\bibinfo
  {journal} {EPL (Europhysics Letters)}\ }\textbf {\bibinfo {volume} {56}},\
  \bibinfo {pages} {319} (\bibinfo {year} {2001})}\BibitemShut {NoStop}%
\bibitem [{\citenamefont {Singh}\ \emph {et~al.}(2012)\citenamefont {Singh},
  \citenamefont {Chatterji}, \citenamefont {Winkler},\ and\ \citenamefont
  {Gompper}}]{Singh_JPCM_2012}%
  \BibitemOpen
  \bibfield  {author} {\bibinfo {author} {\bibfnamefont {S.~P.}\ \bibnamefont
  {Singh}}, \bibinfo {author} {\bibfnamefont {A.}~\bibnamefont {Chatterji}},
  \bibinfo {author} {\bibfnamefont {R.~G.}\ \bibnamefont {Winkler}},\ and\
  \bibinfo {author} {\bibfnamefont {G.}~\bibnamefont {Gompper}},\ }\bibfield
  {title} {\bibinfo {title} {Conformational and dynamical properties of
  ultra-soft colloids in semi-dilute solutions under shear flow},\ }\href@noop
  {} {\bibfield  {journal} {\bibinfo  {journal} {J. Phys.: Condens. Matter}\
  }\textbf {\bibinfo {volume} {24}},\ \bibinfo {pages} {464103} (\bibinfo
  {year} {2012})}\BibitemShut {NoStop}%
\bibitem [{\citenamefont {Lifson}\ and\ \citenamefont
  {Jackson}(1962)}]{lifson1962self}%
  \BibitemOpen
  \bibfield  {author} {\bibinfo {author} {\bibfnamefont {S.}~\bibnamefont
  {Lifson}}\ and\ \bibinfo {author} {\bibfnamefont {J.~L.}\ \bibnamefont
  {Jackson}},\ }\bibfield  {title} {\bibinfo {title} {On the self-diffusion of
  ions in a polyelectrolyte solution},\ }\href@noop {} {\bibfield  {journal}
  {\bibinfo  {journal} {The Journal of Chemical Physics}\ }\textbf {\bibinfo
  {volume} {36}},\ \bibinfo {pages} {2410} (\bibinfo {year}
  {1962})}\BibitemShut {NoStop}%
\bibitem [{\citenamefont {Kim}\ and\ \citenamefont
  {Netz}(2015)}]{kim2015barrier}%
  \BibitemOpen
  \bibfield  {author} {\bibinfo {author} {\bibfnamefont {W.~K.}\ \bibnamefont
  {Kim}}\ and\ \bibinfo {author} {\bibfnamefont {R.~R.}\ \bibnamefont {Netz}},\
  }\bibfield  {title} {\bibinfo {title} {Barrier-induced dielectric counterion
  relaxation at super-low frequencies in salt-free polyelectrolyte solutions},\
  }\href@noop {} {\bibfield  {journal} {\bibinfo  {journal} {The European
  Physical Journal E}\ }\textbf {\bibinfo {volume} {38}},\ \bibinfo {pages} {1}
  (\bibinfo {year} {2015})}\BibitemShut {NoStop}%
\bibitem [{\citenamefont {Fischer}\ and\ \citenamefont
  {Netz}(2013)}]{fischer2013low}%
  \BibitemOpen
  \bibfield  {author} {\bibinfo {author} {\bibfnamefont {S.}~\bibnamefont
  {Fischer}}\ and\ \bibinfo {author} {\bibfnamefont {R.}~\bibnamefont {Netz}},\
  }\bibfield  {title} {\bibinfo {title} {Low-frequency collective exchange mode
  in the dielectric spectrum of salt-free dilute polyelectrolyte solutions},\
  }\href@noop {} {\bibfield  {journal} {\bibinfo  {journal} {The European
  Physical Journal E}\ }\textbf {\bibinfo {volume} {36}},\ \bibinfo {pages} {1}
  (\bibinfo {year} {2013})}\BibitemShut {NoStop}%
\bibitem [{\citenamefont {Heyda}\ and\ \citenamefont
  {Dzubiella}(2012)}]{heyda2012ion}%
  \BibitemOpen
  \bibfield  {author} {\bibinfo {author} {\bibfnamefont {J.}~\bibnamefont
  {Heyda}}\ and\ \bibinfo {author} {\bibfnamefont {J.}~\bibnamefont
  {Dzubiella}},\ }\bibfield  {title} {\bibinfo {title} {Ion-specific counterion
  condensation on charged peptides: Poisson--boltzmann vs. atomistic
  simulations},\ }\href@noop {} {\bibfield  {journal} {\bibinfo  {journal}
  {Soft Matter}\ }\textbf {\bibinfo {volume} {8}},\ \bibinfo {pages} {9338}
  (\bibinfo {year} {2012})}\BibitemShut {NoStop}%
\bibitem [{\citenamefont {Radhakrishnan}\ and\ \citenamefont
  {Singh}(2019{\natexlab{b}})}]{radhakrishnan2019force}%
  \BibitemOpen
  \bibfield  {author} {\bibinfo {author} {\bibfnamefont {K.}~\bibnamefont
  {Radhakrishnan}}\ and\ \bibinfo {author} {\bibfnamefont {S.~P.}\ \bibnamefont
  {Singh}},\ }\bibfield  {title} {\bibinfo {title} {Force driven transition of
  a globular polyelectrolyte},\ }\href@noop {} {\bibfield  {journal} {\bibinfo
  {journal} {The Journal of chemical physics}\ }\textbf {\bibinfo {volume}
  {151}},\ \bibinfo {pages} {174902} (\bibinfo {year}
  {2019}{\natexlab{b}})}\BibitemShut {NoStop}%
\bibitem [{\citenamefont {Raspaud}\ \emph {et~al.}(1998)\citenamefont
  {Raspaud}, \citenamefont {De~La~Cruz}, \citenamefont {Sikorav},\ and\
  \citenamefont {Livolant}}]{raspaud1998precipitation}%
  \BibitemOpen
  \bibfield  {author} {\bibinfo {author} {\bibfnamefont {E.}~\bibnamefont
  {Raspaud}}, \bibinfo {author} {\bibfnamefont {M.~O.}\ \bibnamefont
  {De~La~Cruz}}, \bibinfo {author} {\bibfnamefont {J.-L.}\ \bibnamefont
  {Sikorav}},\ and\ \bibinfo {author} {\bibfnamefont {F.}~\bibnamefont
  {Livolant}},\ }\bibfield  {title} {\bibinfo {title} {Precipitation of dna by
  polyamines: a polyelectrolyte behavior},\ }\href@noop {} {\bibfield
  {journal} {\bibinfo  {journal} {Biophysical journal}\ }\textbf {\bibinfo
  {volume} {74}},\ \bibinfo {pages} {381} (\bibinfo {year} {1998})}\BibitemShut
  {NoStop}%
\bibitem [{\citenamefont {Sabbagh}\ and\ \citenamefont
  {Delsanti}(2000)}]{sabbagh2000solubility}%
  \BibitemOpen
  \bibfield  {author} {\bibinfo {author} {\bibfnamefont {I.}~\bibnamefont
  {Sabbagh}}\ and\ \bibinfo {author} {\bibfnamefont {M.}~\bibnamefont
  {Delsanti}},\ }\bibfield  {title} {\bibinfo {title} {Solubility of highly
  charged anionic polyelectrolytes in presence of multivalent cations: Specific
  interaction effect},\ }\href@noop {} {\bibfield  {journal} {\bibinfo
  {journal} {The European Physical Journal E}\ }\textbf {\bibinfo {volume}
  {1}},\ \bibinfo {pages} {75} (\bibinfo {year} {2000})}\BibitemShut {NoStop}%
\bibitem [{\citenamefont {Wittmer}\ \emph {et~al.}(1995)\citenamefont
  {Wittmer}, \citenamefont {Johner},\ and\ \citenamefont
  {Joanny}}]{wittmer1995precipitation}%
  \BibitemOpen
  \bibfield  {author} {\bibinfo {author} {\bibfnamefont {J.}~\bibnamefont
  {Wittmer}}, \bibinfo {author} {\bibfnamefont {A.}~\bibnamefont {Johner}},\
  and\ \bibinfo {author} {\bibfnamefont {J.}~\bibnamefont {Joanny}},\
  }\bibfield  {title} {\bibinfo {title} {Precipitation of polyelectrolytes in
  the presence of multivalent salts},\ }\href@noop {} {\bibfield  {journal}
  {\bibinfo  {journal} {Journal de Physique II}\ }\textbf {\bibinfo {volume}
  {5}},\ \bibinfo {pages} {635} (\bibinfo {year} {1995})}\BibitemShut {NoStop}%
\bibitem [{\citenamefont {Pelta}\ \emph {et~al.}(1996)\citenamefont {Pelta},
  \citenamefont {Livolant},\ and\ \citenamefont {Sikorav}}]{pelta1996dna}%
  \BibitemOpen
  \bibfield  {author} {\bibinfo {author} {\bibfnamefont {J.}~\bibnamefont
  {Pelta}}, \bibinfo {author} {\bibfnamefont {F.}~\bibnamefont {Livolant}},\
  and\ \bibinfo {author} {\bibfnamefont {J.-L.}\ \bibnamefont {Sikorav}},\
  }\bibfield  {title} {\bibinfo {title} {Dna aggregation induced by polyamines
  and cobalthexamine},\ }\href@noop {} {\bibfield  {journal} {\bibinfo
  {journal} {Journal of Biological Chemistry}\ }\textbf {\bibinfo {volume}
  {271}},\ \bibinfo {pages} {5656} (\bibinfo {year} {1996})}\BibitemShut
  {NoStop}%
\bibitem [{\citenamefont {Delsanti}\ \emph {et~al.}(1994)\citenamefont
  {Delsanti}, \citenamefont {Dalbiez}, \citenamefont {Spalla}, \citenamefont
  {Belloni},\ and\ \citenamefont {Drifford}}]{delsanti1994phase}%
  \BibitemOpen
  \bibfield  {author} {\bibinfo {author} {\bibfnamefont {M.}~\bibnamefont
  {Delsanti}}, \bibinfo {author} {\bibfnamefont {J.}~\bibnamefont {Dalbiez}},
  \bibinfo {author} {\bibfnamefont {O.}~\bibnamefont {Spalla}}, \bibinfo
  {author} {\bibfnamefont {L.}~\bibnamefont {Belloni}},\ and\ \bibinfo {author}
  {\bibfnamefont {M.}~\bibnamefont {Drifford}},\ }\bibfield  {title} {\bibinfo
  {title} {Phase diagram of polyelectrolyte solutions in presence of
  multivalent salts}\ }(\bibinfo  {publisher} {ACS Publications},\ \bibinfo
  {year} {1994})\BibitemShut {NoStop}%
\bibitem [{\citenamefont {Murayama}\ \emph {et~al.}(2003)\citenamefont
  {Murayama}, \citenamefont {Sakamaki},\ and\ \citenamefont
  {Sano}}]{murayama2003elastic}%
  \BibitemOpen
  \bibfield  {author} {\bibinfo {author} {\bibfnamefont {Y.}~\bibnamefont
  {Murayama}}, \bibinfo {author} {\bibfnamefont {Y.}~\bibnamefont {Sakamaki}},\
  and\ \bibinfo {author} {\bibfnamefont {M.}~\bibnamefont {Sano}},\ }\bibfield
  {title} {\bibinfo {title} {Elastic response of single dna molecules exhibits
  a reentrant collapsing transition},\ }\href@noop {} {\bibfield  {journal}
  {\bibinfo  {journal} {Physical review letters}\ }\textbf {\bibinfo {volume}
  {90}},\ \bibinfo {pages} {018102} (\bibinfo {year} {2003})}\BibitemShut
  {NoStop}%
\bibitem [{\citenamefont {Chang}\ and\ \citenamefont
  {Yethiraj}(2003)}]{chang2003brownian}%
  \BibitemOpen
  \bibfield  {author} {\bibinfo {author} {\bibfnamefont {R.}~\bibnamefont
  {Chang}}\ and\ \bibinfo {author} {\bibfnamefont {A.}~\bibnamefont
  {Yethiraj}},\ }\bibfield  {title} {\bibinfo {title} {Brownian dynamics
  simulations of polyelectrolyte solutions with divalent counterions},\
  }\href@noop {} {\bibfield  {journal} {\bibinfo  {journal} {The Journal of
  chemical physics}\ }\textbf {\bibinfo {volume} {118}},\ \bibinfo {pages}
  {11315} (\bibinfo {year} {2003})}\BibitemShut {NoStop}%
\bibitem [{\citenamefont {Huber}(1993)}]{huber1993calcium}%
  \BibitemOpen
  \bibfield  {author} {\bibinfo {author} {\bibfnamefont {K.}~\bibnamefont
  {Huber}},\ }\bibfield  {title} {\bibinfo {title} {Calcium-induced shrinking
  of polyacrylate chains in aqueous solution},\ }\href@noop {} {\bibfield
  {journal} {\bibinfo  {journal} {The Journal of Physical Chemistry}\ }\textbf
  {\bibinfo {volume} {97}},\ \bibinfo {pages} {9825} (\bibinfo {year}
  {1993})}\BibitemShut {NoStop}%
\bibitem [{\citenamefont {Schweins}\ and\ \citenamefont
  {Huber}(2001)}]{schweins2001collapse}%
  \BibitemOpen
  \bibfield  {author} {\bibinfo {author} {\bibfnamefont {R.}~\bibnamefont
  {Schweins}}\ and\ \bibinfo {author} {\bibfnamefont {K.}~\bibnamefont
  {Huber}},\ }\bibfield  {title} {\bibinfo {title} {Collapse of sodium
  polyacrylate chains in calcium salt solutions},\ }\href@noop {} {\bibfield
  {journal} {\bibinfo  {journal} {The European Physical Journal E}\ }\textbf
  {\bibinfo {volume} {5}},\ \bibinfo {pages} {117} (\bibinfo {year}
  {2001})}\BibitemShut {NoStop}%
\bibitem [{\citenamefont {Jia}\ \emph {et~al.}(2012)\citenamefont {Jia},
  \citenamefont {Yang}, \citenamefont {Gong},\ and\ \citenamefont
  {Zhao}}]{jia2012dynamic}%
  \BibitemOpen
  \bibfield  {author} {\bibinfo {author} {\bibfnamefont {P.}~\bibnamefont
  {Jia}}, \bibinfo {author} {\bibfnamefont {Q.}~\bibnamefont {Yang}}, \bibinfo
  {author} {\bibfnamefont {Y.}~\bibnamefont {Gong}},\ and\ \bibinfo {author}
  {\bibfnamefont {J.}~\bibnamefont {Zhao}},\ }\bibfield  {title} {\bibinfo
  {title} {Dynamic exchange of counterions of polystyrene sulfonate},\
  }\href@noop {} {\bibfield  {journal} {\bibinfo  {journal} {The Journal of
  chemical physics}\ }\textbf {\bibinfo {volume} {136}},\ \bibinfo {pages}
  {084904} (\bibinfo {year} {2012})}\BibitemShut {NoStop}%
\bibitem [{\citenamefont {Grass}\ and\ \citenamefont
  {Holm}(2009)}]{grass2009polyelectrolytes}%
  \BibitemOpen
  \bibfield  {author} {\bibinfo {author} {\bibfnamefont {K.}~\bibnamefont
  {Grass}}\ and\ \bibinfo {author} {\bibfnamefont {C.}~\bibnamefont {Holm}},\
  }\bibfield  {title} {\bibinfo {title} {Polyelectrolytes in electric fields:
  Measuring the dynamical effective charge and effective friction},\
  }\href@noop {} {\bibfield  {journal} {\bibinfo  {journal} {Soft Matter}\
  }\textbf {\bibinfo {volume} {5}},\ \bibinfo {pages} {2079} (\bibinfo {year}
  {2009})}\BibitemShut {NoStop}%
\bibitem [{\citenamefont {Kesselheim}\ \emph {et~al.}(2014)\citenamefont
  {Kesselheim}, \citenamefont {M\"uller},\ and\ \citenamefont
  {Holm}}]{Holm_PRL_2014}%
  \BibitemOpen
  \bibfield  {author} {\bibinfo {author} {\bibfnamefont {S.}~\bibnamefont
  {Kesselheim}}, \bibinfo {author} {\bibfnamefont {W.}~\bibnamefont
  {M\"uller}},\ and\ \bibinfo {author} {\bibfnamefont {C.}~\bibnamefont
  {Holm}},\ }\bibfield  {title} {\bibinfo {title} {Origin of current blockades
  in nanopore translocation experiments},\ }\href
  {https://doi.org/10.1103/PhysRevLett.112.018101} {\bibfield  {journal}
  {\bibinfo  {journal} {Phys. Rev. Lett.}\ }\textbf {\bibinfo {volume} {112}},\
  \bibinfo {pages} {018101} (\bibinfo {year} {2014})}\BibitemShut {NoStop}%
\bibitem [{\citenamefont {Kowalczyk}\ and\ \citenamefont
  {Dekker}(2012)}]{Dekker_ACSNANO_2012}%
  \BibitemOpen
  \bibfield  {author} {\bibinfo {author} {\bibfnamefont {S.~W.}\ \bibnamefont
  {Kowalczyk}}\ and\ \bibinfo {author} {\bibfnamefont {C.}~\bibnamefont
  {Dekker}},\ }\bibfield  {title} {\bibinfo {title} {Measurement of the docking
  time of a dna molecule onto a solid-state nanopore},\ }\href@noop {}
  {\bibfield  {journal} {\bibinfo  {journal} {Nano letters}\ }\textbf {\bibinfo
  {volume} {12}},\ \bibinfo {pages} {4159} (\bibinfo {year}
  {2012})}\BibitemShut {NoStop}%
\end{thebibliography}%

 \end{document}